\documentclass[lettersize,journal]{IEEEtran}
\usepackage{amsmath,amsfonts}
\usepackage{algorithmic}
\usepackage{algorithm}
\usepackage{array}
\usepackage[caption=false,font=normalsize,labelfont=sf,textfont=sf]{subfig}
\usepackage{textcomp}
\usepackage{stfloats}
\usepackage{url}
\usepackage{verbatim}
\usepackage{graphicx}
\usepackage{cite}
\usepackage{bm}

\usepackage{color}

\begin{document}

\title{A Different View of Sigma-Delta Modulators Under the Lens of Pulse Frequency Modulation}

\author {Victor~Medina,~\IEEEmembership{Graduate Student Member,~IEEE,}
         Pieter~Rombouts,~\IEEEmembership{Senior Member,~IEEE,} 
        Luis~Hernandez,~\IEEEmembership{Senior Member,~IEEE}
\thanks{V. Medina and L. Hernandez are with the Department of Electronic Technology, Carlos III University, Madrid,
Spain, (e-mail: luish@ing.uc3m.es).}
\thanks{P. Rombouts is with ELIS, Ghent University, Belgium.}
\thanks{© 2024 IEEE.  Personal use of this material is permitted.  Permission from IEEE must be obtained for all other uses, in any current or future media, including reprinting/republishing this material for advertising or promotional purposes, creating new collective works, for resale or redistribution to servers or lists, or reuse of any copyrighted component of this work in other works. DOI: 10.1109/MCAS.2024.3352552}
}

\markboth{Journal of \LaTeX\ Class Files,~Vol.~14, No.~8, August~2021}%
{Shell \MakeLowercase{\textit{et al.}}: A Sample Article Using IEEEtran.cls for IEEE Journals}

\maketitle

\IEEEpeerreviewmaketitle

\begin{abstract}
The fact that VCO-ADCs produce noise-shaped quantization noise suggests that a link between frequency modulation and Sigma-Delta modulation should exist. The connection between a VCO-ADC and a first-order Sigma-Delta modulator has been already explained using Pulse Frequency Modulation. In this paper, we attempt to extend this explanation to a generic Sigma-Delta modulator. We show that the link between Sigma-Delta modulation and Pulse Frequency Modulation relies in a sampling invariance property that defines the equivalence between both entities. This equivalence property, allows to go beyond the white quantization noise model of a Sigma-Delta modulator, revealing the origin of some nonlinear phenomena. We first predict spurious tones which cannot be explained by circuit non linearity. Multi-bit and single-bit modulators are shown to belong to a same generic class of systems. Finally, quantizer overload is analyzed using our model. The results are applied to Continuous-Time Sigma-Delta modulators of orders one, two and three and then extended to a generic case.
\end{abstract}


\section{Introduction}
\IEEEPARstart{I}{n} the '80s decade of last century, the world faced the advent of digital audio and telephony. One of the main technological problems at that time, was the lack of high resolution data converters that were cheap and suitable for mass production. Resistor ladder DACs and successive approximation ADCs were among the few available data converter chips, and they required laser trimming to achieve CD audio quality. Trimming was a costly production step that contributed to push digital audio to the most expensive market segments. By the end of the decade, Sigma-Delta modulators came as a solution to popularize communications and multimedia products. As a consequence, and despite being first described in the '60s \cite{Brahm, inose}, it was not until then that Sigma-Delta modulators became a hot topic, focusing many researchers in their analysis \cite{candy}.

Since the beginning, Sigma-Delta modulation was very useful implemented as switched capacitor ADCs or purely digital DACs but a mystery from the system level point of view. Many attempts were made to explain its spectra even in the simplest form \cite{gray}, the stability of Sigma-Delta modulators \cite{schreierconvex,feely} or its anomalous dynamic range \cite{ardalan}. However, the only simplified analysis that remains after the years is the white quantization noise analysis model \cite{candy, libro_candy} complemented with simulations to prove stability \cite{schreierempirical}. An in depth understanding of the quantization noise and operation of Sigma-Delta modulators has not been achieved despite the years. 

In the opinion of the authors, this is due to the fact that system level understanding and hardware implementations have followed opposite ways. First Sigma-Delta modulators were discrete systems implemented with switched capacitors (ADCs) or digital circuits (DACs). Then, development of Continuous-Time Sigma-Delta (CTSD) modulators showed that all Discrete-Time Sigma-Delta (DTSD) modulators are a discretization in the sampling instants of an equivalent CTSD modulator \cite{schreier_ct}. Afterwards, Voltage Controlled Oscillators (VCO) were proposed as an alternative implementation of Sigma-Delta modulators \cite{hovin} or as replacements of the quantizer and last integrator of a CTSD modulator \cite{Perrott_Straayer}. 

VCO-based ADCs, share the same noise shaping properties of CTSD modulators, showing first-order noise shaping in its basic form \cite{colorines}. In\cite{zierhofer2008frequency}, the links between Pulse Frequency Modulation (PFM) and first-order Sigma-Delta modulation were first envisioned. Later on, a detailed analysis of VCO-based ADCs \cite{Hernandez2015}, established an analytical closed form for the DFT of a first-order noise shaped sequence generated by a VCO-ADC. This analysis is based in considering the VCO-ADC as a PFM encoder followed by a square pulse shaping filter and a sampler. The square pulse filter introduces periodical nulls in the output spectrum at multiples of the sampling frequency. After the sampler, noise shaping properties are given by the aliases of the modulation  \cite{Gutierrez2018}. It is interesting to observe how this aliasing mechanism can produce exactly the same output sequence of a first-order CTSD modulator. This fact suggests that the analysis based on PFM theory could be extended to high order CTSD modulators \cite{medina_iscas} and predict their properties to some extent without resorting to the white-noise linear model. Therefore, the path to an extended analysis of Sigma-Delta modulation links first PFM with CTSD modulation and afterwards DTSD modulators are explained as a particular implementation of CTSD modulators, which is just the opposite direction in which mainstream technology advanced over the years. 

This paper is an attempt to do an extension of VCO-ADC analyses based on PFM to CTSD modulation. We observe the similarities between a generic CTSD modulator and an oscillator, once the sampler is replaced by a monostable pulse generator. We show that a CTSD modulator, which includes an integrator and a sampler in a feedback loop, is identical to a PFM modulator whose oscillation is sampled outside of the oscillation loop. This sampling invariance property is a key element to our analysis because, given a generic Sigma-Delta modulator, it allows to propose an alternative system (the PFM equivalent) which inherits all the properties of the original modulator but dissagregates the quantization error into some PFM side bands and a pseudo-random sampling alias component.  

The paper is structured as follows: Section II reviews the basic concepts of PFM. Section III shows the equivalence of a PFM modulator and a first-order CTSD modulator even if the quantizer is multi-bit, giving a rigorous mathematical proof of the sampling invariance property in Appendix A. Afterwards, the invariance property is applied to second and third-order CTSD modulator examples. The results of Section III are extended to any Sigma-Delta modulator with a Cascade of Integrators with distributed Feedback (CIFB) structure \cite{Schreier2017understanding} in Annex B. The goal of Section IV is to explain Sigma-Delta modulation beyond the white quantization noise model, showing the root cause of some nonlinear phenomena that cannot be attributed to circuit non-idealities. In section V, we address the conditions under which the PFM model is valid and its implication with the quantizer overload of a Sigma-Delta modulator. In Section VI we draw some practical conclusions and propose new research topics. 

\section{Review of PFM modulation}

\subsection{PFM definition}
\begin{figure}[t]
	\centering
	\includegraphics[width=\columnwidth,keepaspectratio]{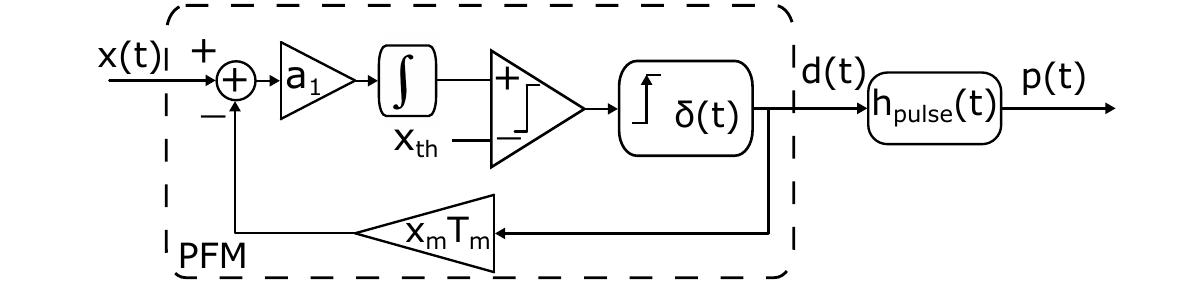}
 	\caption{Model of a PFM modulator.} 
    \label{fig:fig1_NW}
\end{figure}
 The system shown in Fig.~\ref{fig:fig1_NW} corresponds to the standard definition of a PFM modulator as described in, for example, \cite{Bayly1968,Fitch1947}. It is built around an ideal integrator with gain $a_1$ followed by a comparator with threshold $x_{th} \geq 0$. The rising edges of the comparator output, trigger a block generating Dirac delta impulses, which are fed with negative feedback to the input of the integrator through a gain $x_m\cdot T_m$. For positive input signals $x(t)>0$, the comparator will trigger Dirac delta impulses in $d(t)$. We can write an equation describing the time instants $t_k$ where the Dirac delta impulses are located \footnote{We assume the Heaviside function $u(t)=\int _{-\infty} ^{t}\delta(\tau)d\tau$ to be one in $t=0$, $u(0)=1$.}:

\begin{eqnarray}
d(t)=\sum _{k=0} ^\infty \delta(t-t_k)  \\ 
t_k \, \mid \, 0=a_1\cdot\int_{t_{k-1}} ^{t_k} (x(t)-x_m \cdot T_m \cdot \delta(t-t_{k}))\,dt  \nonumber \\
t_k \, \mid \, x_m \cdot T_m=\int_{t_{k-1}} ^{t_k} x(t)\,dt
\label{eq:pfm_recurrence}
\end{eqnarray}

Note that if the input was always negative, the integrator output would never reach the threshold and the system would stop. The train of impulses $d(t)$ gets through a pulse shaping filter with impulse response $h_{pulse}(t)$ to generate $p(t)$.  Although theoretically, the pulse shaping filter may be any filter, in practice a square pulse is selected to have a digital-friendly two-level signal. The impulse response $h_{pulse}(t)$ and transfer function $H_{pulse}(s)$ of this square shaping pulse are defined as follows, 
\footnote{In this work, we use lowercase letters for time domain signal representations, while uppercase letters are used for transform domain (either Fourier, s or Z-domain) representations.}
considering $u(t)$ the Heaviside unit step function:
\begin{equation}
  h_{pulse}(t) = (u(t)-u(t-T_m)) \label{eq:1stOrderPulseShaperExplicit}
\end{equation}
\begin{equation}
  H_{pulse}(s) = \frac{1-e^{-sT_m}}{s} \label{eq:1stOrderSincExplicit}
\end{equation}
Thus, this pulse shaping filter, turns the Dirac delta impulses in $d(t)$ into a series of pulses of fixed length $T_{m}$ and unit amplitude in $p(t)$.  

If we assume that $x(t)$ is a DC signal $x_{DC}$, then $d(t)$ will be a periodic signal whose frequency depends linearly on $x_{DC}$ through a constant $k_d$. To calculate $k_d$, we can equate the area of one feedback delta impulse with the integral in one period $T$ of $d(t)$ for $x_{DC}$:

\begin{eqnarray}
0=a_1\cdot\int_{t_{k-1}} ^{t_{k-1}+T} (x_{DC}-x_m \cdot T_m \cdot \delta(t-t_{k-1}-T))\,dt  \nonumber \\ \nonumber
f= 1/T = x_{DC}/(x_{m}\cdot T_m)  \nonumber \\
k_{d}=1/(x_{m} \cdot T_m) 
\label{eq:pfm_dc_input}
\end{eqnarray}

According to \eqref{eq:pfm_dc_input}, if we apply a midrange input signal $x_{DC}=x_{m}/2$, the output frequency will be $f_0=(2T_m)^{-1}$. However, and as a difference to a practical hardware PFM modulator, the system of Fig.~\ref{fig:fig1_NW} does not require the input signal to be upper bounded and could theoretically encode an input $x(t)$ between 0 and $\infty$ given that the feedback is composed of infinitely narrow Dirac delta pulses. For instance, if the input $x_{DC}$ grows by an integer multiple $M$ of $x_{m}/2$, the modulator will produce a frequency proportionally larger:
 \begin{eqnarray}
 x_{DC}=M \cdot x_m/2
  \\
  f_0=M/2T_m    \nonumber
 	\label{eq:Mf0}
 \end{eqnarray}

We will use this property in the following sections to model multi-bit quantizers. We can write the frequency dependence of the PFM modulator with the input $x(t)$ as follows:
 \begin{eqnarray}
 	f(t)=k_{d}\cdot x(t),  0<x(t)<\infty
 	\label{eq:sect1:fosc}
 \end{eqnarray}

\begin{figure}[t]
	\centering
	\includegraphics[width=\columnwidth,keepaspectratio]{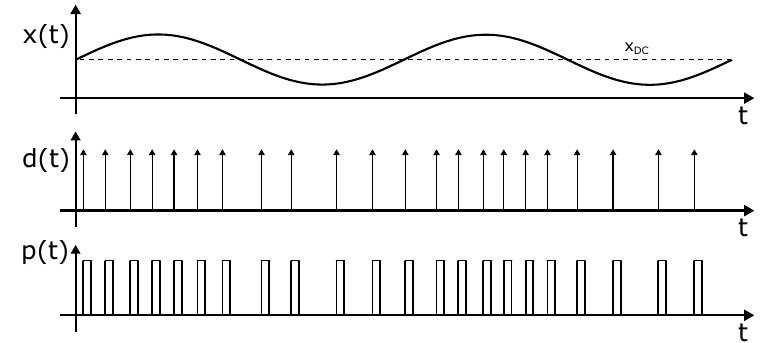}
	\caption{Input signal, PFM and pulse shaping filter output.} \label{fig:fig_input_dt_pt}
\end{figure}

Intuitively and as a consequence of \eqref{eq:sect1:fosc}, if the input signal $x(t)$ changes over time, the average frequency of the modulator output $d(t)$ will be proportional to the input signal. Fig.~\ref{fig:fig_input_dt_pt} shows the time domain representation of signals $x(t)$, $d(t)$ and $p(t)$ in an example where $x(t)$ has a DC component $x_{DC}=x_m/2$ and a sinusoidal AC component. 
In the next subsection we will review the spectral contents of $d(t)$ for sinusoidal inputs.
\subsection{Spectral properties of PFM signals}
The mathematical description of the spectrum of PFM signals has been a question of practical importance in communications, control theory, biology and power electronics, to name a few fields. One of the first publications solving this question for sinusoidal inputs is \cite{Fitch1947} for square pulses and \cite{Bayly1968} for Dirac delta streams. We are particularly interested in the results of \cite{Bayly1968} because the premises for the derivation of the calculations and the definition of PFM in \cite{Bayly1968} match exactly with the proposed reference modulator of Fig.~\ref{fig:fig1_NW} and the integral equation \eqref{eq:pfm_recurrence}. If we consider a sinusoidal input signal with a DC component $x_m/2$, amplitude $A<x_m/2$ and frequency $f_x$, then $x(t)$ becomes:
\begin{eqnarray}
x(t)=x_m/2+A\cdot \cos(2\pi f_xt).
\end{eqnarray}
Signal $d(t)$ can then be expanded into a trigonometric series as follows \cite{Bayly1968}:
\begin{eqnarray}
d(t)=f_0+k_d\cdot A \cdot \cos(2\pi f_xt) + m(t),\nonumber \\ \nonumber m(t)=2f_0\sum_{q=1}^{\infty}\sum_{r=-\infty}^{\infty}J_r\left ( q\frac{A\cdot k_d}{f_x} \right ) \\ \cdot \left ( 1+\frac{rf_x}{qf_0} \right )\cos(2\pi (qf_0+rf_x)t)
\label{eq:dt_spectrm}
\end{eqnarray}
where $J_r(\cdot)$ is the Bessel function of the first kind. In this equation we can distinguish two terms, one representing the same input signal multiplied by the modulator gain $k_d$ plus a DC offset due to $f_0$ and another term $m(t)$ representing some modulation side bands centered at $f_0$ and its harmonics. 
\begin{figure}[t]
\center
 \includegraphics[width=\columnwidth,keepaspectratio] {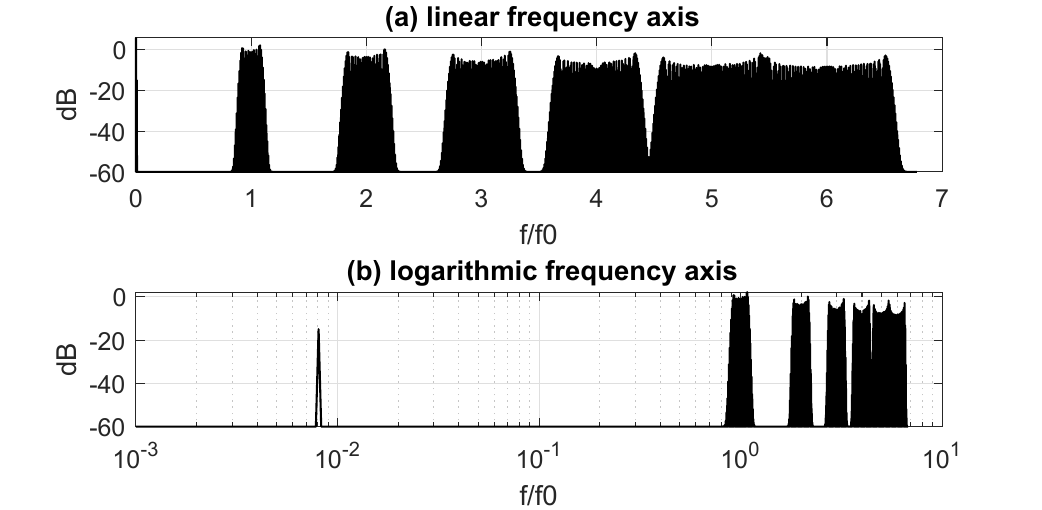}
\caption{Calculated spectra of $d(t)$ with the first six side bands for the case of an input signal with A=1/8 and $f_0=128f_x$, (a) linear frequency axis, (b) logarithmic frequency axis, x(t) can be observed at $f_x=1/128$ with an amplitude of -19 dB corresponding to A=1/8.} \label{fig:NSidebandsSpectrum}
\end{figure}
The spectral distribution of the modulation side bands depends in a complicated way on the signal parameters. In general each modulation side band, roughly consists of a relatively pronounced central lobe and a tail that decays relatively rapidly. This is particularly the case when the input signal frequency $f_x$ is small relative to the rest oscillation frequency $f_0$. This situation is shown in Fig.~\ref{fig:NSidebandsSpectrum}(a) and (b) in an example. It is clear from the figure that the bulk of every side band’s energy is contained in the central lobe which has a limited bandwidth. If we denote $BW_q$ the value of this limited bandwidth for the $q^{th}$ side band, we found that the following relationship approximately holds \cite{carson}:
\begin{eqnarray}
BW_q\approx 2qAk_{d}.
\label{eq:pfm_bw}
\end{eqnarray}
The higher order side bands become wider with index $q$, which has the consequence that the spectra of these higher order side bands in Fig.~\ref{fig:NSidebandsSpectrum} overlap. Due to this, the high frequency spectrum resembles white noise.

\section{The Connection between Sigma-Delta Modulation and PFM: Sampling Invariance} \label{section_equivalence}

As VCO-ADCs developed, it was clear that both first-order Sigma-Delta modulators and VCO-ADCs produced first-order noise shaped sequences \cite{hovin}. In \cite{Hernandez2015} it was shown that a sampled PFM modulator produces exactly the same sequence as a VCO followed by a XOR-based frequency to digital decoder. It may seem that VCO-ADCs and first-order CTSD modulators share a first-order noise shaped spectrum on the average, however, we are going to show in this section that they are exactly the same system when the reference model for the PFM modulator is used. The practical consequence of this is that we may place a sampler inside of a first-order loop (like in a first-order CTSD modulator) or outside of the loop (like in the reference PFM modulator) and get two identical sequences. This fact will be referred as the \textit{sampling invariance property of first-order CTSD modulators} and will be a key development in explaining the operation of higher order CTSD modulators and the structure of their quantization noise.

\subsection{First Order Case}
\begin{figure}[t]
	\centering
	\includegraphics[width=1\columnwidth,keepaspectratio]{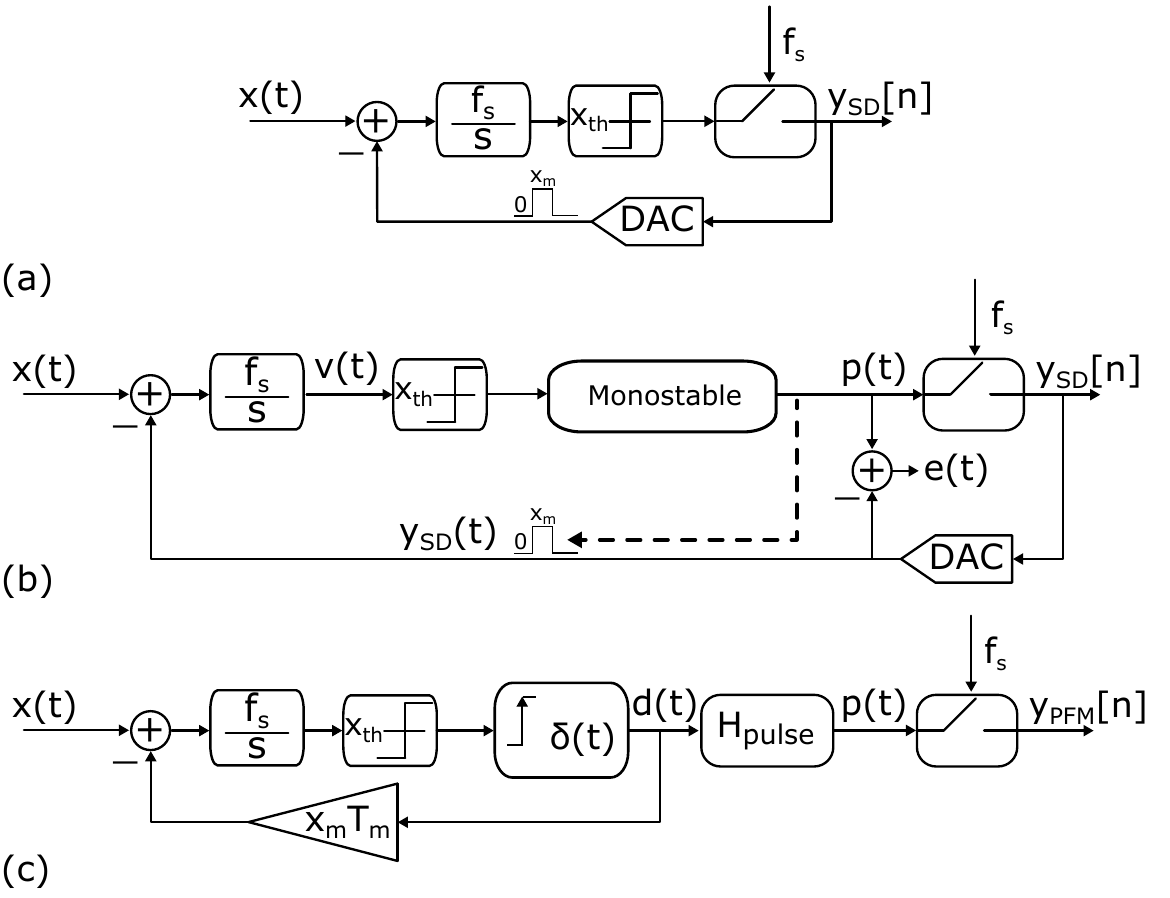}
  \caption{Equivalence between a first-order single-bit CTSD and a PFM based model.}
	 \label{fig:fig1_VM}
\end{figure}

\begin{figure}[t]
	\centering
	\includegraphics[width=1\columnwidth,keepaspectratio]{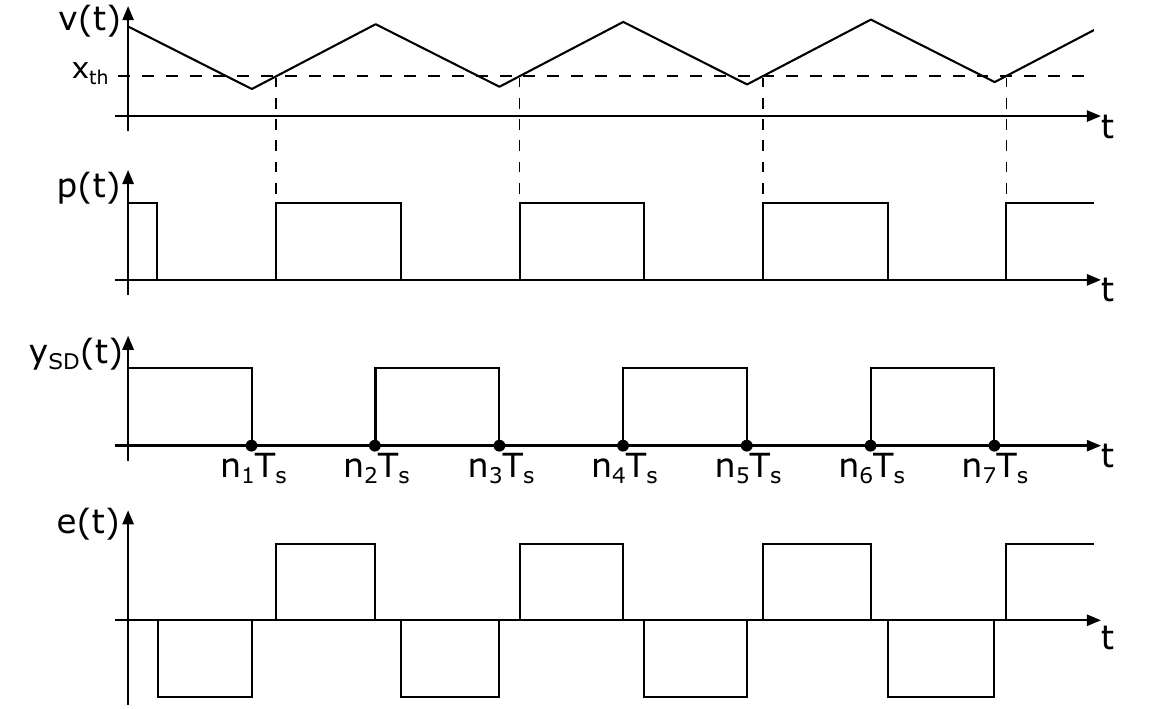}
  \caption{Time domain waveforms of $v(t)$, $p(t)$, $y_{SD}(t)$ and $e(t)$ of Fig.~\ref{fig:fig1_VM}}
	 \label{fig:impinv_example}
\end{figure}

 Fig.~\ref{fig:fig1_VM}(a) shows a first-order, single-bit CTSD modulator. To ease the definition of the equivalence with the PFM modulator described in Section II, we will assign values to some parameters. The quantizer threshold of Fig.~\ref{fig:fig1_VM}(a) will be $x_{th}=1/2$, we will define $x_m=1$, the feedback DAC will produce a NRZ pulse defined between 0 and $x_{m}=1$ and the input signal $x(t)$ will be bounded between 0 and $x_{m}=1$. Finally we will set parameter $T_m$ to be equal to a sampling period defined as $T_m=T_s=1/f_s$. These constraints represent a DC shift and a scaling over the standard single-bit Sigma-Delta modulator with outputs $+1$ and $-1$ to operate always with positive signals defined between 0 and $+1$, and does not incur in a loss of generality. The system of Fig.~\ref{fig:fig1_VM}(a) can be transformed in the equivalent system depicted in Fig.~\ref{fig:fig1_VM}(b), which is described in detail in \cite{medina_iscas}, to give us some intuition on the sampling invariance principle. We have illustrated the operation of this system with the example of Fig.~\ref{fig:impinv_example}. In the system of Fig.~\ref{fig:fig1_VM}(b), the threshold crossings of the integrator output $v(t)$, trigger the generation of square pulses in a monostable \cite{medina_iscas} producing signal $p(t)$, (see Fig.~\ref{fig:impinv_example}). These pulses are sampled with a digital sampler resulting in signal $y_{SD}(t)$ and re-injected in the loop through a single-bit DAC. The sampling operation, aligns the pulses in $p(t)$ to occur at the sampling instants $nT_s$. Therefore, the sampler introduces an aliasing error $e(t)$ in $p(t)$ which has variable width square pulses that average to a zero mean. Signal $p(t)$ resembles the output of a pulse frequency modulator, which is sampled afterwards like in a VCO-ADC, with the difference that the monostable is embedded in the loop. Surprisingly, a time domain simulation shows that the sampled sequence generated by this system is identical regardless that the feedback is connected to $y_{SD}(t)$ or to $p(t)$ (see dashed line in Fig.~\ref{fig:fig1_VM}(b)) for DC inputs. 
 In Fig.~\ref{fig:fig1_VM}(c) we have reproduced the standard PFM modulator depicted in Fig.~\ref{fig:fig1_NW} followed by a sampler. As a difference to Fig.~\ref{fig:fig1_VM}(b), the loop has Dirac delta impulses as feedback and the monostable is replaced by filter $h_{pulse}$. 
 
\begin{figure}
	\centering
	\includegraphics[width=\columnwidth,keepaspectratio]{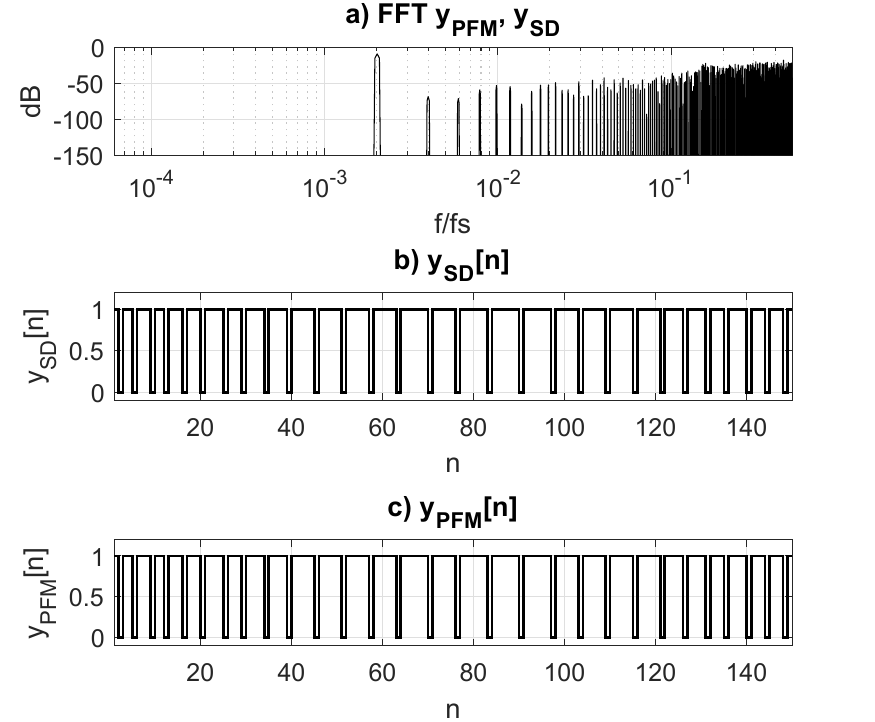}
	\caption{Signal equivalences between a CTSD modulator and its PFM equivalent.}
\label{fig:PFM_CTSD_signals}
\end{figure}

 In Fig.~\ref{fig:PFM_CTSD_signals} we have compared the behaviour of the conventional CTSD modulator of Fig.~\ref{fig:fig1_VM}(a) and Fig.~\ref{fig:fig1_VM}(c) with a time domain simulation of the same sinusoidal input signal. The simulation shows the sample-by-sample coincidence of $y_{PFM}$ and $y_{SD}$. Given that both systems, Fig.~\ref{fig:fig1_VM}(a) and Fig.~\ref{fig:fig1_VM}(c) produce the same output for the same input we can infer that they are equivalent.

The explanation of the system equivalence illustrated in Fig.~\ref{fig:fig1_VM} is not rigorous since the monostable of Fig.~\ref{fig:fig1_VM}(b) has some restrictions (see \cite{medina_iscas}), which are difficult to model mathematically, and because it is based on an empirical coincidence. However, in Appendix A we show a mathematically rigorous proof of the equivalence between the systems of Fig.~\ref{fig:fig1_VM}(a) and Fig.~\ref{fig:fig1_VM}(c) using the complete induction method. The advantage of the PFM equivalent of Fig.~\ref{fig:fig1_VM}(c) is that the sampler is outside of the loop and that the behavior of this system has been studied in \cite{Gutierrez2018}, which permits to decompose the quantization noise of the white noise CTSD model into a more insightful way.

\begin{figure}[t]
	\centering
	\includegraphics[width=\columnwidth,keepaspectratio]{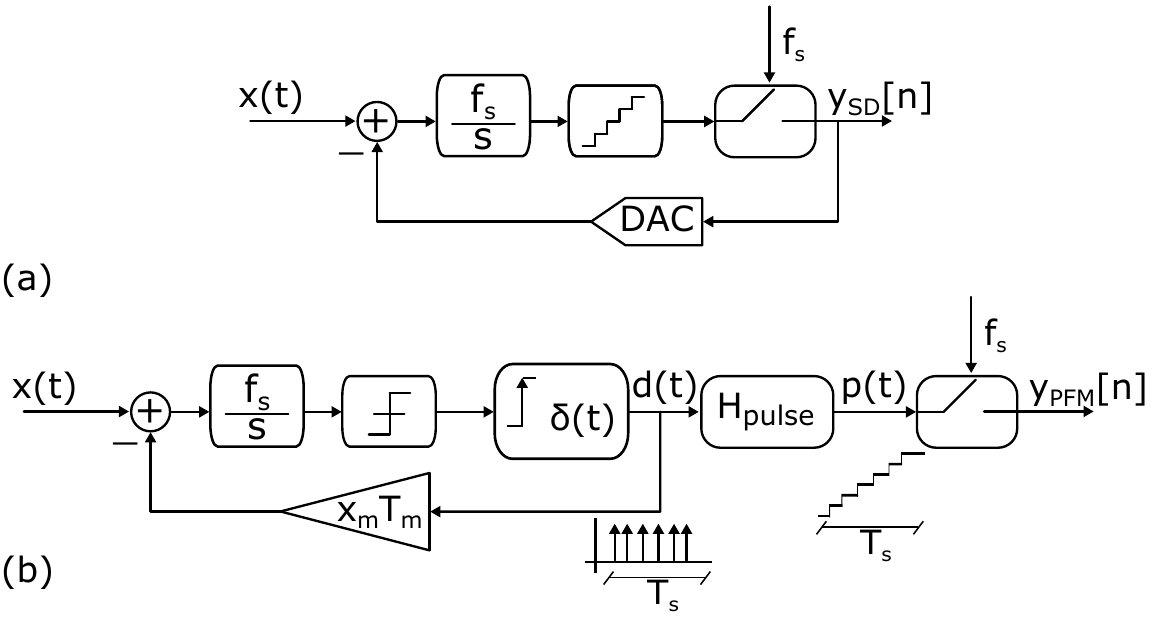}
	\caption{Equivalence between a first-order multi-bit CTSD and a PFM based model.} 
    \label{fig:fig2_VM}
\end{figure}

\begin{figure}
	\centering
	\includegraphics[width=\columnwidth,keepaspectratio]{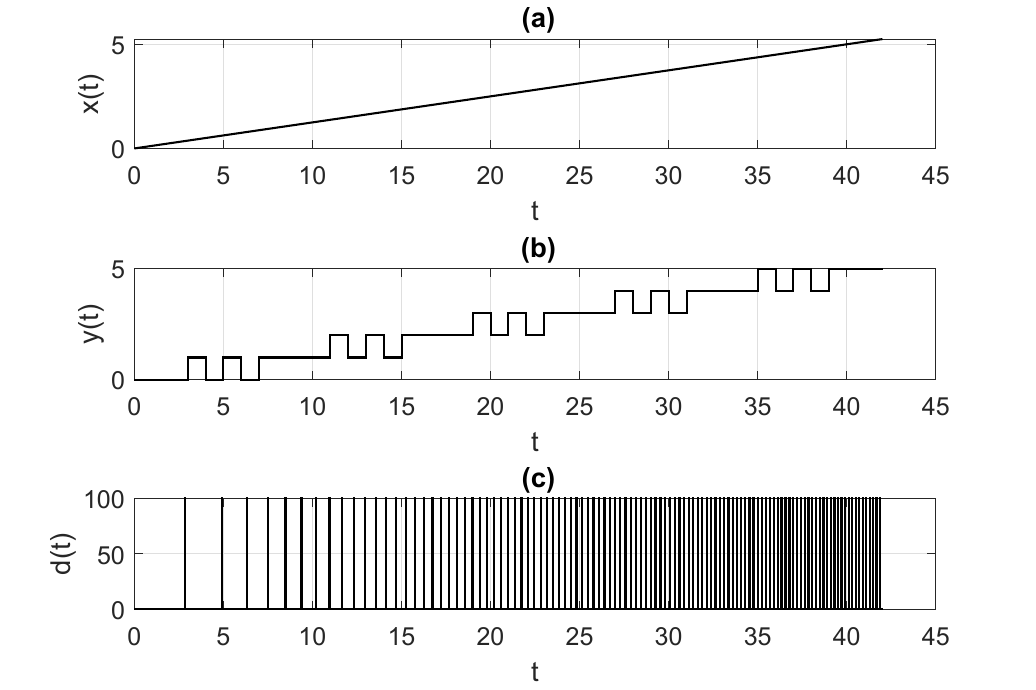}
	\caption{Multi-bit behaviour in a PFM modulator configured as a first-order CTSD modulator.}
\label{fig:pfm_multibit_proof}
\end{figure}

It is also possible to build a multi-bit Sigma-Delta modulator based on the PFM equivalent model. The relationship between quantization and frequency modulation was inferred previously in a different context in \cite{quantized_sine_zierhofer,quantized_sine_claasen}. We now will consider that the input signal $x(t)$ is only lower bounded, $0 \leq x(t)$. Fig.~\ref{fig:fig2_VM}(a) depicts a multi-bit first-order CTSD modulator where the quantization step size is equal to 1 and the integrator gain is equal to $f_s$. Its equivalent PFM model is displayed in Fig.~\ref{fig:fig2_VM}(b). The difference here is that the oscillation frequency of the PFM modulator is allowed to go beyond the sampling frequency $f_s$. As the input signal becomes greater than 1, there will be delta pulses spaced less than $T_s$ and the output $p(t)$ will present more than 2 levels due to the overlap of several $h_{pulse}$ functions, that is, the encoder will have a multi-bit behavior. Thus, for the described first-order CTSD modulator, the difference between single-bit and multi-bit lies on the magnitude of $x(t)$ or conversely in the ratio between $f_s$ and the PFM oscillation frequency. Fig.~\ref{fig:pfm_multibit_proof} shows an example of this multi-bit behaviour. A slow ramp is applied to the system of Fig.~\ref{fig:fig2_VM}(b) as $0<x(t)<6$ (see Fig.~\ref{fig:pfm_multibit_proof}(a)). The output of the PFM equivalent system ($y(t)$ in Fig.~\ref{fig:pfm_multibit_proof}(b)) displays a multilevel signal following the input. Signal $y(t)$ matches exactly with the output of the equivalent CTSD multi-bit modulator. The PFM output $d(t)$ is shown as a delta train in Fig.~\ref{fig:pfm_multibit_proof}(c).

\begin{figure}[t]
	\centering
	\includegraphics[width=\columnwidth,keepaspectratio]{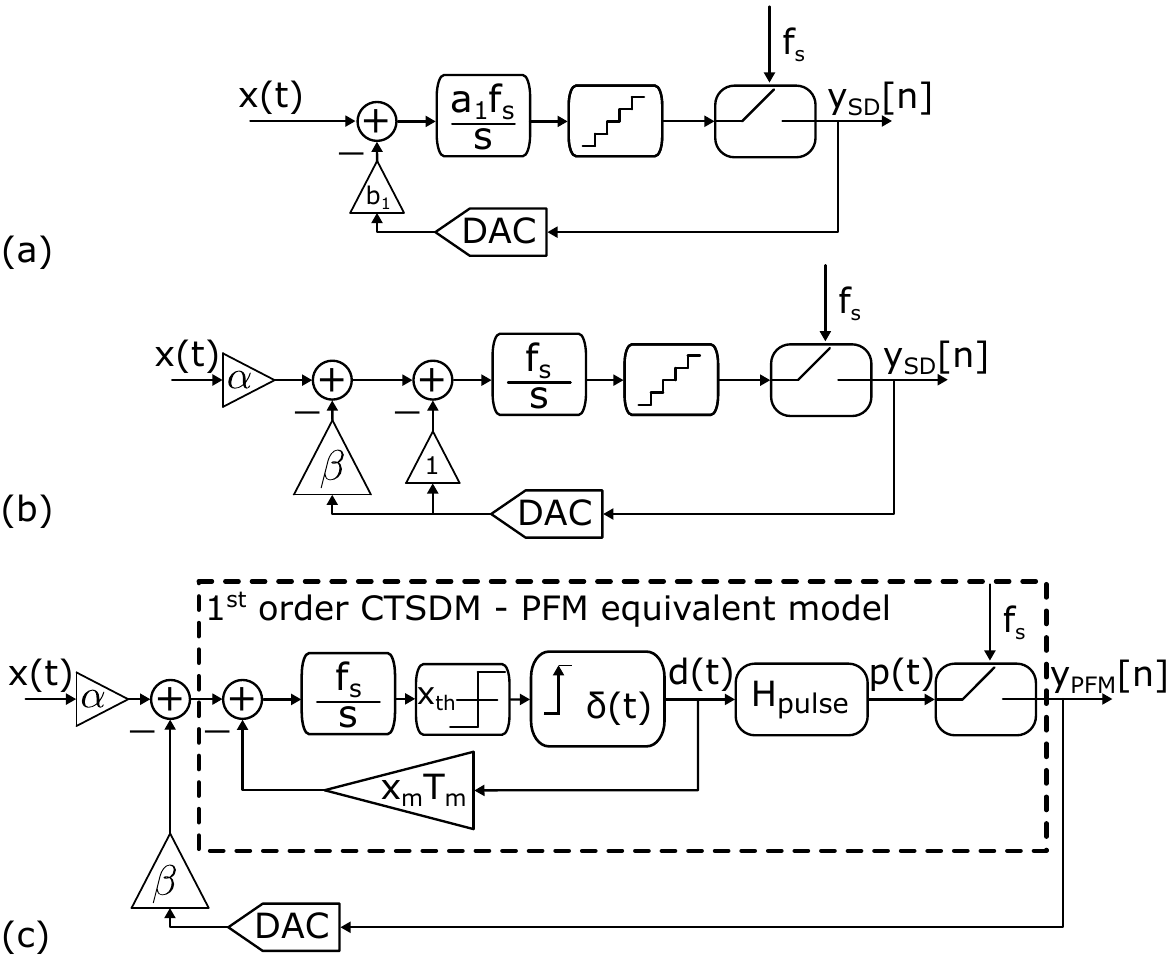}
	\caption{General equivalence between (a) first-order multi-bit CTSD, (b) scaled first-order multi-bit CTSD and (c) PFM based model.}
 \label{fig:fig_generalEquivalence_1stOrder}
\end{figure}

An important remark must be made about the gain of the integrator in Fig. \ref{fig:fig1_VM}(a). We will assume next that the gain of the integrator is scaled by $a_1 \cdot f_s$. At system level, in any single-bit CTSD modulator, the gain of the last integrator is irrelevant to the output sequence. This is due to the nature of its quantizer, which basically works as a comparator. This makes the first-order single-bit case a very special one. Furthermore, if we change the NRZ pulse by any arbitrary pulse shape of equivalent area, the modulator will remain unaffected \cite{pavan_method_moments}.
In the single-bit case and while $0 \leq x(t) \leq 1$, the gain $a_1$ of systems of Fig.~\ref{fig:fig1_VM}(a) and Fig.~\ref{fig:fig1_VM}(c) can change in both to an arbitrary value, producing identical results. This can be proven by observing that in \eqref{eq:pfm_dc_input}, coefficient $a_1$ does not influence $k_d$. 
However, when the quantizer uses more than one level, the gain of the integrator does influence the behavior of the system. Fig.~\ref{fig:fig_generalEquivalence_1stOrder}(a) shows a more general case of a first-order Sigma-Delta modulator, where the integrator and the feedback gains are scaled by coefficients $a_1$ and $b_1$ respectively. To obtain the PFM equivalent, we first transform the system of Fig.~\ref{fig:fig_generalEquivalence_1stOrder}(a) to a system where the the last loop is composed by the canonical first-order CTSD modulator shown in Fig.~\ref{fig:fig2_VM}(a), that is, with unity feedback gain, and an integrator gain equal to $f_s$. This transformation is shown in Fig.~\ref{fig:fig_generalEquivalence_1stOrder}(b). For both systems to be equivalent, coefficients $\beta$ and $\alpha$ must guarantee that Fig.~\ref{fig:fig_generalEquivalence_1stOrder}(b) and Fig.~\ref{fig:fig_generalEquivalence_1stOrder}(a) have the same open loop impulse response with respect to the DAC and the input. We now can replace this canonical first-order CTSD modulator by the PFM modulator. Consequently, this transformation requires adding a loop around the PFM modulator. The equivalent PFM model is shown in Fig.~\ref{fig:fig_generalEquivalence_1stOrder}(c). For this first-order modulator, the values of $\alpha$ and $\beta$ can be computed as follows:
\begin{eqnarray}
\alpha=a_1 , \beta=a_1 \cdot b_1 -1
\label{eq:alphabet}
\end{eqnarray}
For higher order modulators, the same procedure will be used to obtain the values of $\alpha$ and $\beta$.

\subsection{Second Order Case} 
\begin{figure}[t]
	\centering
	\includegraphics[width=\columnwidth,keepaspectratio]{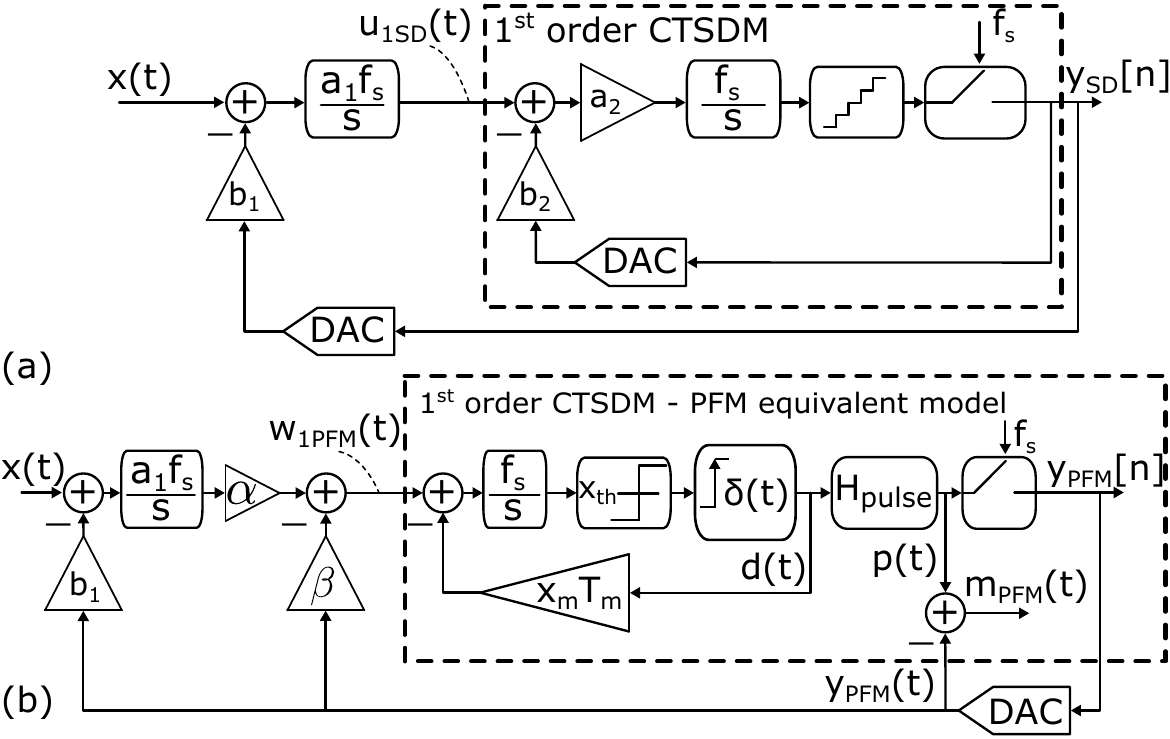}
	\caption{General equivalence for a second-order multi-bit CTSD modulator.} \label{fig:pfm_equivalence_2nd_general}
\end{figure}

We can also build a PFM equivalent of a standard second-order CTSD modulator, as the one shown in Fig.~\ref{fig:pfm_equivalence_2nd_general}(a). The two integrator loops in Fig.~\ref{fig:pfm_equivalence_2nd_general}(a) have been drawn nested to point out the resemblance of the inner loop (surrounded by a dashed line) with a first-order Sigma-Delta modulator. A possible approach to define a PFM equivalent for the modulator of Fig.~\ref{fig:pfm_equivalence_2nd_general}(a) is to replace this inner loop by the generic first-order PFM equivalent modulator of Fig.~\ref{fig:fig_generalEquivalence_1stOrder}(c). This replacement is shown in Fig.~\ref{fig:pfm_equivalence_2nd_general}(b), where the sampler has been placed outside of the PFM modulator loop. In this case, the values of parameters $\alpha$ and $\beta$ can be defined as follows:
\begin{eqnarray}
\alpha=a_2 , \beta=a_2 \cdot b_2 -1
\label{eq:alphabet2}
\end{eqnarray}

\begin{figure}
	\centering
	\includegraphics[width=\columnwidth,keepaspectratio]{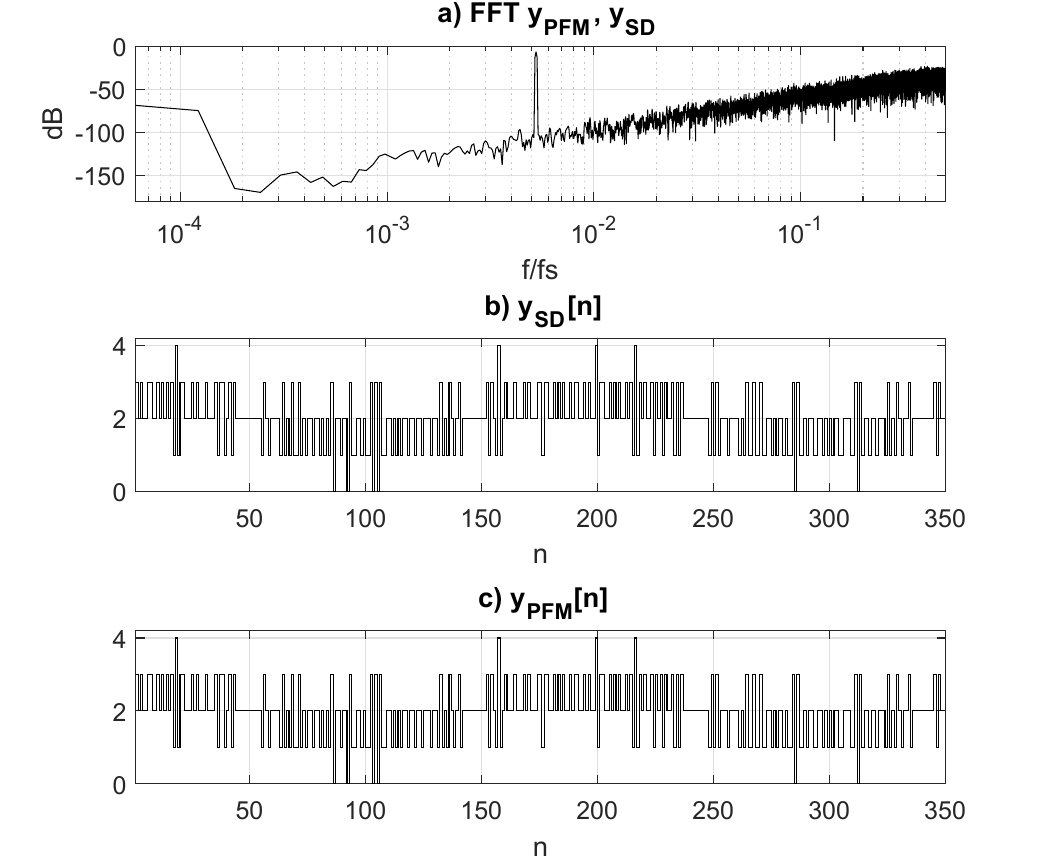}
	\caption{Comparison between a multi-bit second-order CTSD modulator and its PFM equivalent: (a) FFT of both systems, (b) Time domain of $y_{SD}$, (c) Time domain of $y_{PFM}$.} \label{fig:sine_simulation_2nd}
\end{figure}

Fig.~\ref{fig:sine_simulation_2nd}(a) shows the FFT of a simulation of a multi-bit version of both systems with $x_m=1$ and $x_{DC}=2$, which define a 5 level quantizer (levels 0 to 4). Loop coefficients are $a_1=1$, $a_2=1$, $b_1=1$ and $b_2=3/2$. As a consequence, $\alpha=1$ and $\beta=1/2$. The results for a $-6.5~\text{dB}_{\text{FS}}$ sinusoidal input are shown in Fig.~\ref{fig:sine_simulation_2nd}(b) and Fig.~\ref{fig:sine_simulation_2nd}(c), which display identical output sequences. 

\begin{figure}
	\centering
	\includegraphics[width=\columnwidth,keepaspectratio]{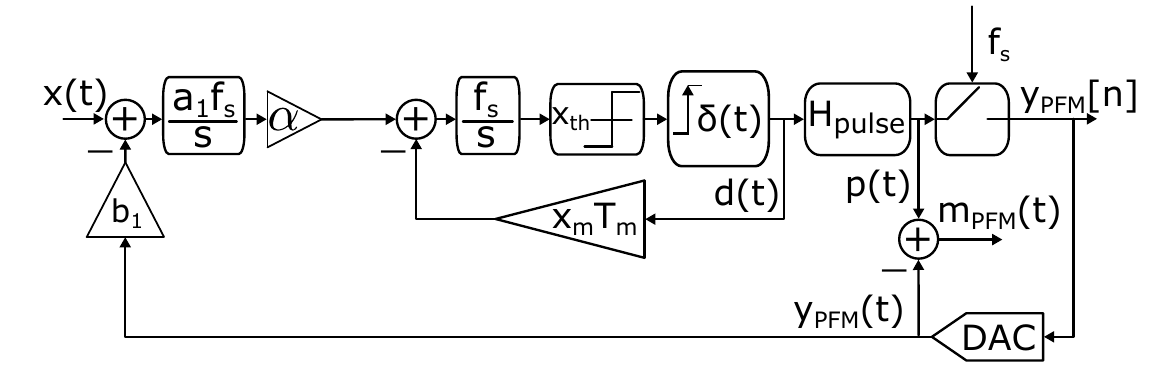}
	\caption{PFM equivalence for a second-order single-bit  CTSD modulator.} \label{fig:pfm_equivalence_2nd_single_bit}
\end{figure}

The scaling proposed in \eqref{eq:alphabet}, \eqref{eq:alphabet2} also depends on the magnitude of the state variable driving the PFM modulator, which in some single-bit modulators, may not be valid. A clear example of this situation happens in the second-order single-bit CTSD modulator. The single-bit version of the second-order CTSD modulator cannot be scaled like the multi-bit modulator of Fig.~\ref{fig:pfm_equivalence_2nd_general} because it would not produce a single-bit output except for very small input amplitudes. State variable $u_{1SD}(t)$ (see Fig.~\ref{fig:pfm_equivalence_2nd_general}) is, on average, larger than the input signal, which requires to scale properly the PFM modulator to remain single-bit even for large amplitudes. We propose the model depicted in Fig.~\ref{fig:pfm_equivalence_2nd_single_bit}, where we add a coefficient $\alpha$ in order to scale the input to the PFM as follows:
\begin{eqnarray}
 \alpha= 1/b_2
 \label{eq:alphabet_single_bit}
\end{eqnarray}

Coefficient $a_2$ does not appear in our model as its value is irrelevant in a single-bit modulator. In Section V we will discuss in detail the implications of this scaling in the quantizer overload of CTSD modulators.

\begin{figure}
	\centering
	\includegraphics[width=\columnwidth,keepaspectratio]{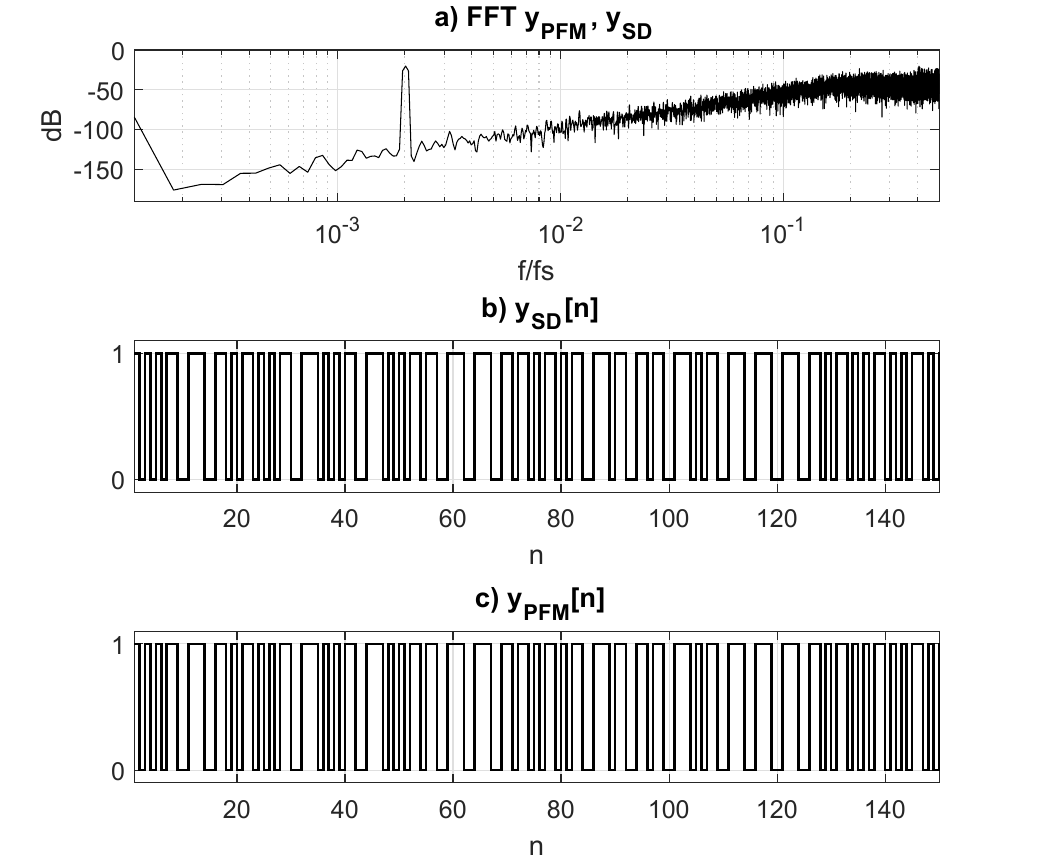}
	\caption{Comparison between single-bit second-order CTSD modulator and its PFM equivalent: (a) FFT of both systems, (b) Time domain of $y_{SD}$, (c) Time domain of $y_{PFM}$.}
 \label{fig:figure_2nd_single_bit}
\end{figure}

In Fig.~\ref{fig:figure_2nd_single_bit}(a) we have represented the FFT of a simulation of a single-bit second-order Sigma-Delta modulator and its PFM equivalent system, 
with the standard coefficients ($a_1=a_2=b_1=1, b_2=3/2, x_m=1$). Fig.~\ref{fig:figure_2nd_single_bit}(b) and Fig.~\ref{fig:figure_2nd_single_bit}(c) show the time domain identical outputs for both models.

\subsection{Third Order Example and Extension to a Generic Modulator}
\begin{figure}[t]
	\centering
	\includegraphics[width=\columnwidth,keepaspectratio]{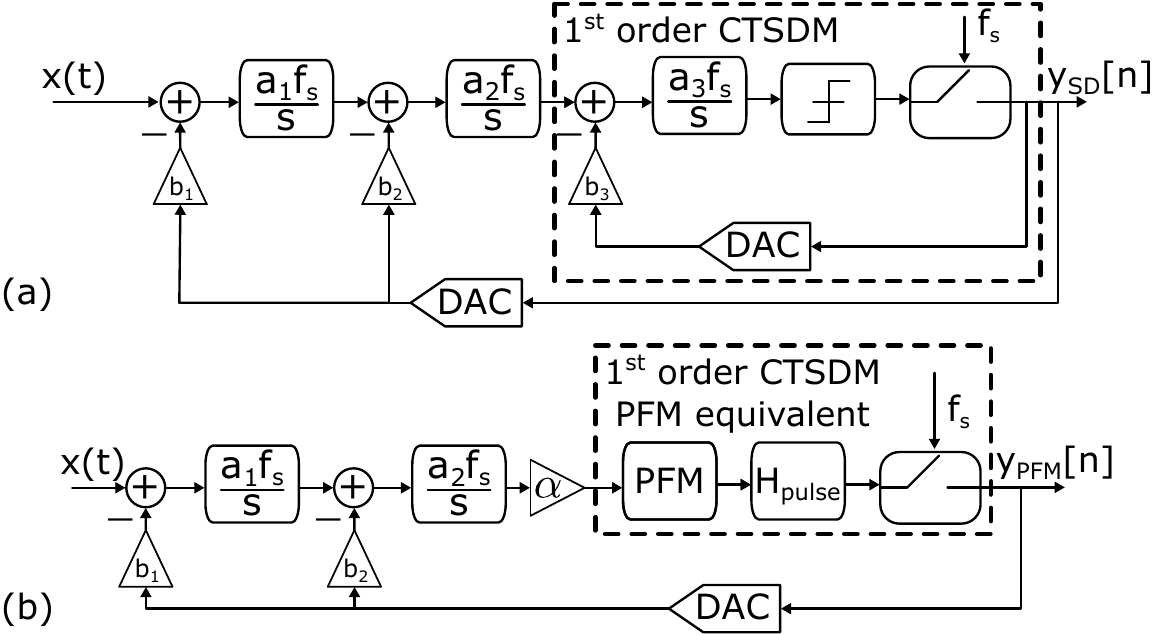}
	\caption{Third order CTSD modulator reproduced from \cite{ortmanns_thirdOrder_singleBit} (a), and its equivalent PFM model (b).} \label{fig:thirdOrder_Ortmanns_models}
\end{figure}

\begin{figure}[t]
	\centering
	\includegraphics[width=\columnwidth,keepaspectratio]{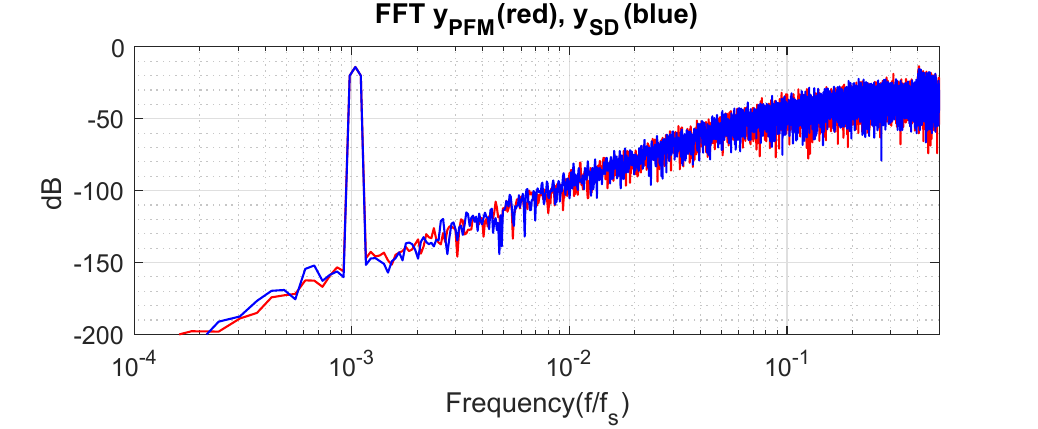}
	\caption{FFT for systems of Fig.~\ref{fig:thirdOrder_Ortmanns_models}.} \label{fig:thirdOrder_Ortmanns_FFT_CTDS_PFM}
\end{figure}

We can also build a PFM equivalent for a low-pass third-order CTSD modulator. As an example, we have chosen the single-bit Sigma-Delta modulator described in \cite{ortmanns_thirdOrder_singleBit}. Fig.~\ref{fig:thirdOrder_Ortmanns_models}(a) reproduces the original system \cite{ortmanns_thirdOrder_singleBit}, where $b_{1}=0.05$, $b_{2}=0.3$ and $b_{3}=0.641$ and all integrator gains are equal to $f_s$ ($a_1=a_2=a_3=1$). Fig.~\ref{fig:thirdOrder_Ortmanns_models}(b) shows the PFM equivalent, obtained using the same transformation applied to the second-order modulator. In \cite{ortmanns_thirdOrder_singleBit}, the designers scaled state variables to reach the peak Signal-to-Quantization Noise Ratio (SQNR) close to full scale. A benefit of this scaling is that the PFM modulator in the equivalent system gets a small input. As a consequence, we can replicate the scaling shown in either Fig.~\ref{fig:pfm_equivalence_2nd_general} or Fig.~\ref{fig:pfm_equivalence_2nd_single_bit} and yet keep the single-bit behaviour in both cases. In Fig.~\ref{fig:thirdOrder_Ortmanns_models}(b), the PFM modulator has been simplified as a single block representing the PFM loop of Fig.~\ref{fig:fig1_VM}(c). For the PFM equivalent, the following scaling values of the parameters are used: $\alpha=a_3$ and $\beta=a_3 \cdot b_3-1$. Fig.~\ref{fig:thirdOrder_Ortmanns_FFT_CTDS_PFM} shows two FFTs for both systems outputs in Fig.~\ref{fig:thirdOrder_Ortmanns_models}. Initial conditions were intentionally selected such that $y_{SD}$ and $y_{PFM}$ don't match sample by sample and consequently their FFTs differ one from the other. However, both outputs match sample by sample for adequate initial conditions. 

\begin{figure}[t]
	\centering
	\includegraphics[width=\columnwidth,keepaspectratio]{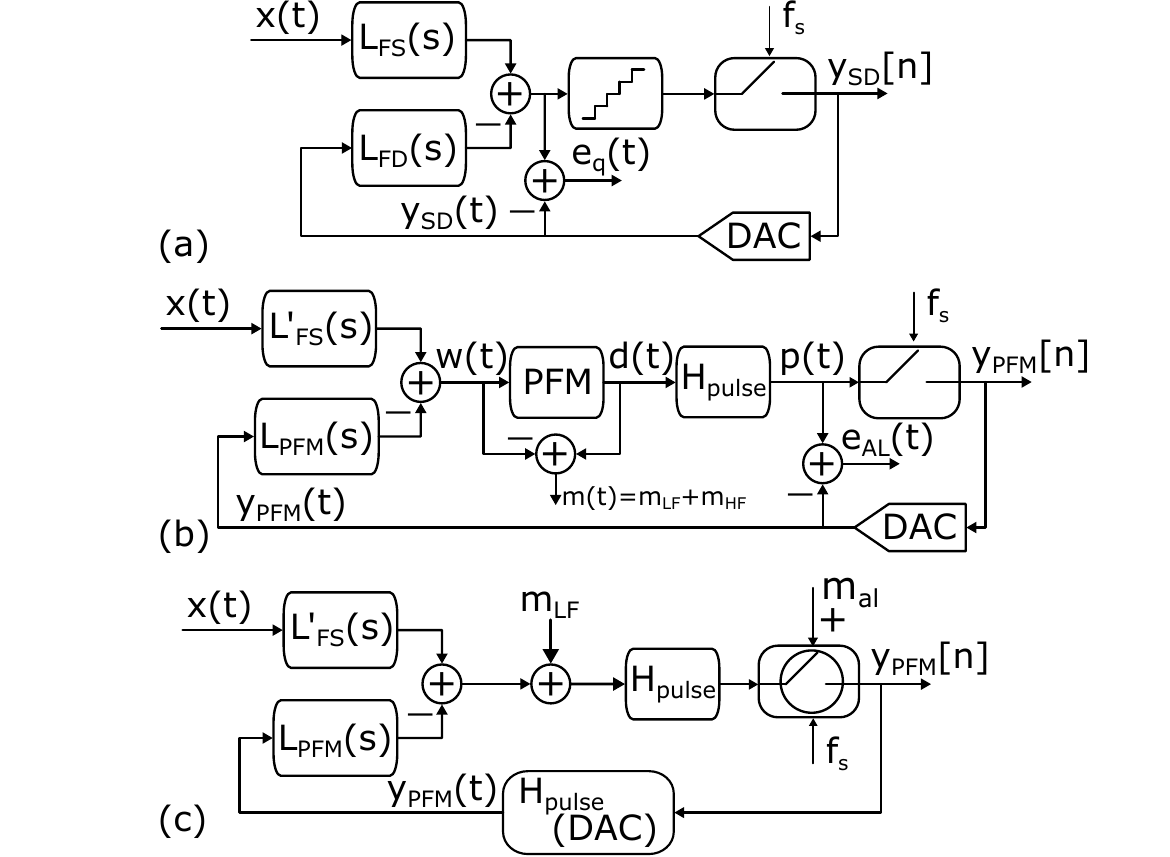}
	\caption{Error source comparision in a CTSD modulator: (a) White noise model, (b) Generic PFM model and (c) Signal processing equivalent.} \label{fig:noise_source_comparison}
\end{figure}

The case of the second and third-order modulators allows us to propose a method to represent a generic CTSD modulator with a multiple feedback structure. Fig.~\ref{fig:noise_source_comparison}(a) shows the white noise model of a CTSD modulator described by two transfer functions, $L_{FS}(s)$ for the input and $L_{FD}(s)$ for the feedback loop. The modulator produces the output sequence $y_{SD}[n]$ and we consider quantization noise $e_q(t)$ to be added at the quantizer. In Fig.~\ref{fig:noise_source_comparison}(b) we have represented the equivalent PFM model, which would produce an identical output sequence $y_{PFM}[n]=y_{SD}[n]$ and is split into a different transfer function $L_{PFM}(s)$ for the feedback loop and an input transfer function $L_{FS}'(s)$ modified to reflect the new summation point of input and feedback. The loop contains also a PFM modulator producing a delta pulse stream $d(t)$, a square shaping pulse filter $h_{pulse}(t)$ and a sampler that samples the discrete-amplitude signal $p(t)$. In Appendix B, we propose a methodology to extend the PFM equivalent to any CIFB modulator without local loops in the loop filter and we detail the calculation of $L_{FS}'(s)$.

\section{Explaining Sigma-Delta Modulation beyond the white noise model}

In the previous section, the PFM equivalent of a CTSD modulator has been described. However, this equivalence would be of little interest if it did not bring any further insight into Sigma-Delta modulation. In this section, we are going to show how the PFM equivalent of a CTSD modulator allows to describe the quantization noise beyond the commonly used white noise assumption with the help of Fig.~\ref{fig:noise_source_comparison}. In Fig.~\ref{fig:noise_source_comparison}(b), we can apply our knowledge of the PFM spectral properties expressed by \eqref{eq:dt_spectrm}. The otherwise considered "quantization noise" $e_{q}(t)$ of  Fig.~\ref{fig:noise_source_comparison}(a) is represented by two components in Fig.~\ref{fig:noise_source_comparison}(b). On one hand, we have the PFM output that can be described by a component linearly depending on the input signal combined with modulation side bands caused by the PFM modulation $m(t)$. For simplicity, we will assume that the PFM modulator has unity gain for $w(t)$. The modulation side bands are shown in the sinusoidal case of \eqref{eq:dt_spectrm} centered at multiplies of the rest frequency of the PFM modulator $f_0$. In Fig.~\ref{fig:noise_source_comparison}(b), signal $m(t)$ can be split into the spectral components of the PFM modulator in the Nyquist zone $m_{LF}(t)$, (between 0 and $f_s/2$) and the rest of the modulation side bands at higher frequencies $m_{HF}(t)$. We can write:
\begin{equation}
  d(t) = w(t) + m_{LF}(t) + m_{HF}(t)
\end{equation}
On the other hand, we will have the aliasing error $e_{al}(t)$ introduced by the sample and hold after sampling the discrete level signal $p(t)$. In this section we will use this alternative representation to describe the mechanisms that produce spurious components in Sigma-Delta modulators.

 \begin{figure}[t]
	\centering
	\includegraphics[width=\columnwidth,keepaspectratio]{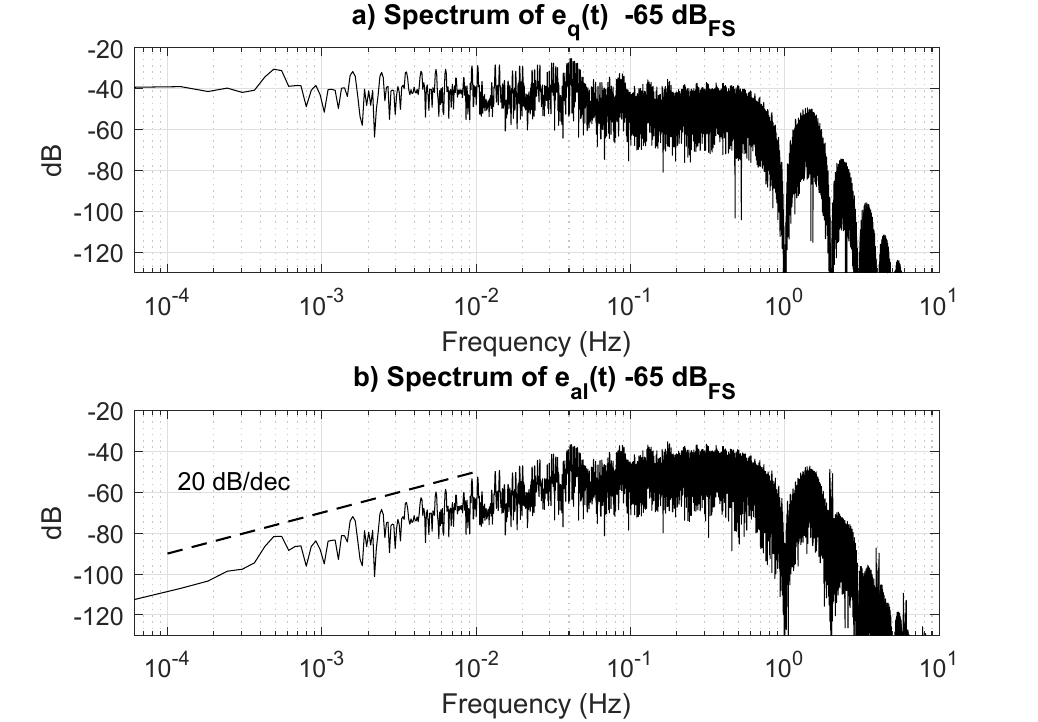}
	\caption{Noise components in (a) CTSD modulator with $-65~\text{dB}_{\text{FS}}$ input, (b) Sampling error in the equivalent PFM model.} \label{fig:noise_components65_part1}
\end{figure}

 \begin{figure}[t]
	\centering
	\includegraphics[width=\columnwidth,keepaspectratio]{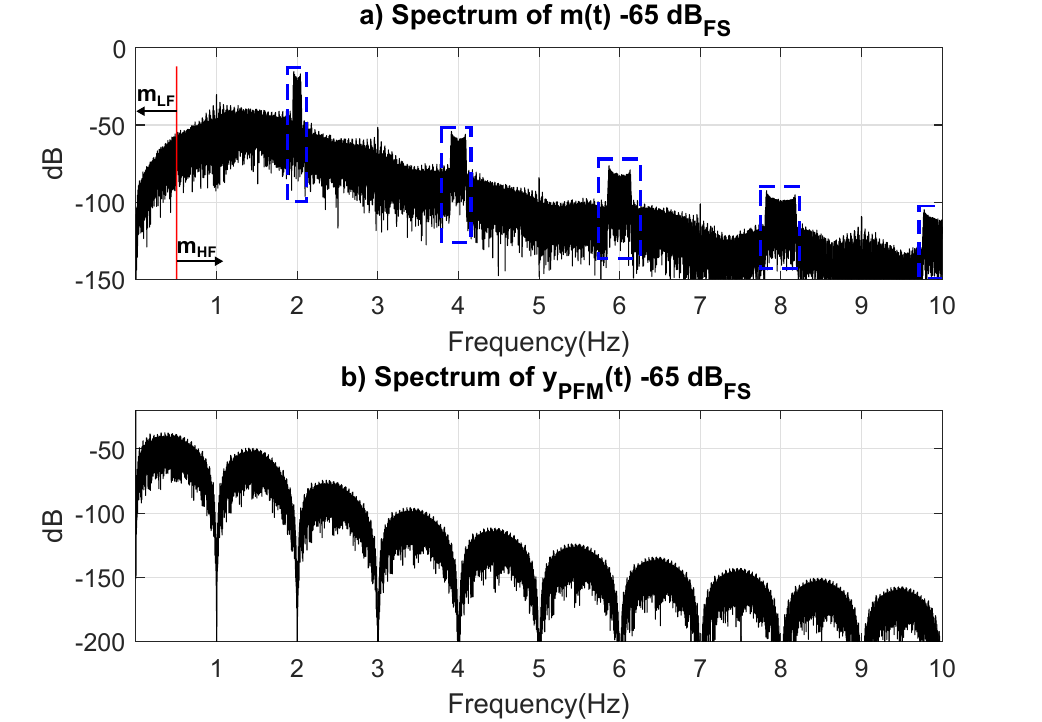}
	\caption{(a) Modulation side bands for a  $-65~\text{dB}_{\text{FS}}$ input, (b) Analog output $y_{PFM}(t)$.} \label{fig:noise_components65_part2}
\end{figure}

To gain some insight in this error decomposition, we have plotted in Fig.~\ref{fig:noise_components65_part1} the spectral contents of each component in the second-order multi-bit example of Fig.~\ref{fig:pfm_equivalence_2nd_general}. In the simulation we have used a 5 level quantizer around a DC value of $x_{DC}=2$ (i.e. $0 \leq x(t) \leq 4$) and a sampling frequency $f_s=1$. 

To visualize spectral contents beyond the sampling frequency, the plots show spectral estimations of the continuous-time signals, assuming that the DAC and sampler implement a zero-order-hold function. In a first simulation, we have used an input tone of $-65~\text{dB}_{\text{FS}}$. Fig.~\ref{fig:noise_components65_part1}(a) shows the spectrum of the quantization noise $e_q(t)$ of the CTSD modulator of Fig.~\ref{fig:noise_source_comparison}(a), showing an approximately uniform noise spectral density, as predicted by the white noise model. Fig.~\ref{fig:noise_components65_part1}(b) shows the spectrum of the aliasing error $e_{al}(t)$ introduced by the sampler of Fig.~\ref{fig:noise_source_comparison}(b) in log-frequency scale, displaying first-order spectral shaping with a slope of 20 dB/dec. Observing the signal flow in Fig.~\ref{fig:noise_source_comparison}(b) we can see that from $w(t)$ to $y_{PFM}(t)$, we have the same model of a first-order Sigma-Delta modulator described in \cite{Hernandez2015}, where a detailed explanation of the first-order shaped nature of $e_{al}(t)$ is given. A summary of the explanation in \cite{Hernandez2015} follows. The aliasing error signal, $e_{al}(t)$ is the difference between $p(t)$ and its sampled version $y_{PFM}(t)$. Now, $p(t)$ is the output of the pulse shaping filter $h_{pulse}(t)$, (see \eqref{eq:1stOrderPulseShaperExplicit}) which has a zero at the sampling frequency $f_s$ and at all integer multiples of $f_s$. Hence the spectrum of  $p(t)$ must have a zero at the sampling frequency $f_s$ (and its integer multiples). After sampling, these components near integer multiples of $f_s$ with their zero, will alias to DC. Therefore, the aliasing error $e_{al}(t)$ must also have one zero at DC. Also, an intuitive explanation for this fact can be drawn from Fig.~\ref{fig:impinv_example} because $e_{al}(t)$ is the difference between a signal composed of square pulses with duration $T_s$ and the same signal sampled with exactly a period $T_s$ \cite{colorines}. Note that this fact holds for all CTSD modulators expressed in its PFM equivalent, and regardless of its order and loop filter transfer function.

Fig.~\ref{fig:noise_components65_part2}(a) shows the spectrum of $m(t)$, the modulation side bands introduced by the PFM modulator. Given that $x_{DC}=2$, $T_m=T_s=1$, and according to \eqref{eq:Mf0}, these modulation side bands are located at the rest oscillation frequency of the PFM modulator $f_0=2f_s$ and its integer multiples. The modulation side bands have been marked with blue dashed boxes in Fig.~\ref{fig:noise_components65_part2}(a). If we observe these modulation side bands, we can see that they are composed of groups of tones following the frequency distribution described by \eqref{eq:dt_spectrm}. However, $m(t)$ also contains spectrally shaped noise components. We have marked in Fig.~\ref{fig:noise_components65_part2}(a) the boundary between the first Nyquist zone (from 0 to $f_s/2$) with a red line to separate components $M_{LF}$ and $M_{HF}$, located above $f_s/2$. Finally, Fig.~\ref{fig:noise_components65_part2}(b) shows the spectrum of $y_{PFM}(t)$, the analog waveform reproduced by the feedback DAC. As could be expected from a sampled signal, it consists of periodic copies of the samples of $p(t)$, containing all aliases from $m(t)$ and weighted in frequency by the sinc function associated with the NRZ DAC. Although it may seem surprising, function $h_{pulse}(t)$ in our model, appears twice in the loop. One $h_{pulse}(t)$ is located before the sampler, which stems from the integration in the last integrator of the loop filter (embedded in the PFM modulator), and another in the feedback DAC. 

 \begin{figure}[t]
	\centering
	\includegraphics[width=\columnwidth,keepaspectratio]{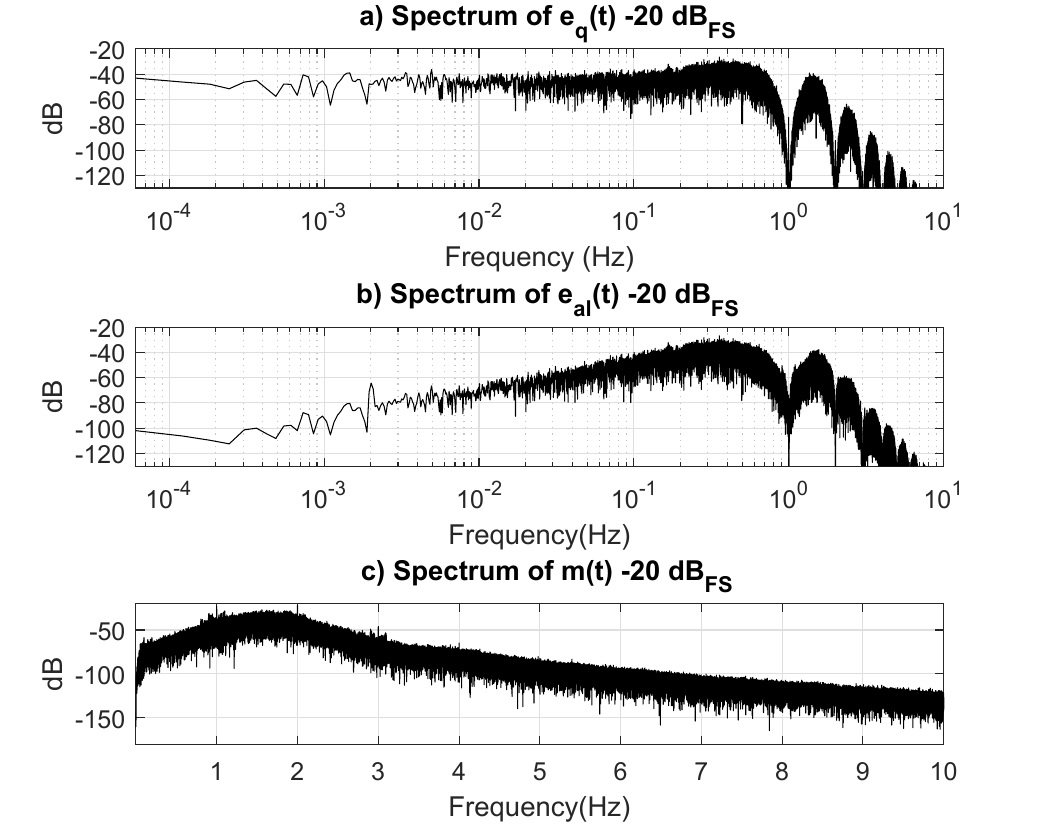}
	\caption{Noise components in (a) CTSD modulator white noise with a $-20~\text{dB}_{\text{FS}}$, (b) Sampling error of PFM equivalent model, (c) Modulation side bands.} \label{fig:noise_components20}
\end{figure}

In general, CTSD modulators are well described by the white noise model when modulation side bands in the spectrum of $m(t)$ are wide and overlap resembling a white noise, which depends on the harmonic order and input amplitude \eqref{eq:pfm_bw}. For instance, if we apply an input signal of $-20~\text{dB}_{\text{FS}}$ instead of $-65~\text{dB}_{\text{FS}}$ in the simulation of Fig.~\ref{fig:noise_components65_part2}, we obtain the plot of Fig.~\ref{fig:noise_components20}, where the modulation side bands are not individually identifiable in the plot, but a continuum resembling white noise. However, there are times this situation does not hold and side bands resemble discrete tones that couple to the signal band due to several mechanisms. Single-bit modulators and modulators with very small inputs are prone to these situations and therefore to show tonal behaviours, not attributable to circuit non-linearity. 

 \begin{figure}[t]
	\centering
	\includegraphics[width=\columnwidth,keepaspectratio]{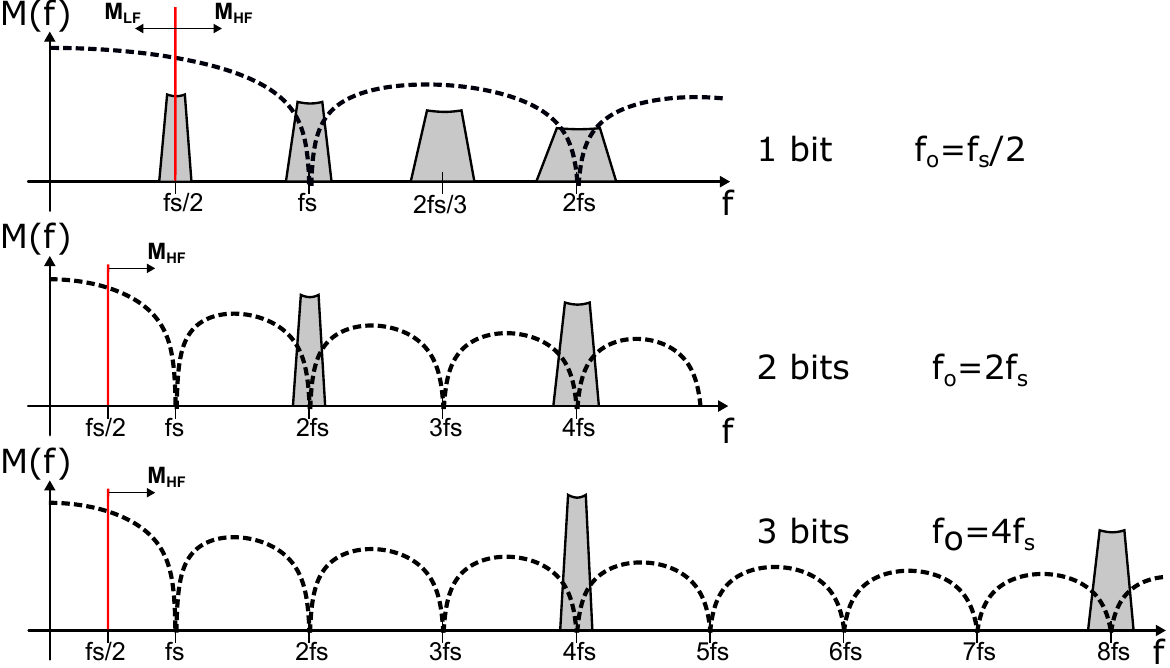}
	\caption{Influence of quantizer resolution in PFM modulation side band distribution. } \label{fig:sidebands_vs_quantizer_size}
\end{figure}

After the empirical observations of Fig.~\ref{fig:noise_components65_part1}, \ref{fig:noise_components65_part2} and Fig.~\ref{fig:noise_components20}, we are going to describe how noise shaping happens in the PFM model of a Sigma-Delta modulator. First we will describe the structure of the modulation side bands in $M(f)$, the spectrum of $m(t)$, which depends on the quantizer bit resolution N. To give rise to $2^N$ different levels at the shaping pulse filter $h_{pulse}(t)$, we need the rest frequency of the PFM modulator to be $2^N$ times faster than that of the single-bit case $f_0=f_s/2$. In our model, this is achieved by simply increasing $x_{DC}$. Fig.~\ref{fig:sidebands_vs_quantizer_size} shows the cases N = 1 bit to N = 3 bit, with the sinc function associated to $H_{pulse}(f)$ superimposed (dotted lines). In the single-bit case (N=1), we still have a modulation side band centered at $f_s/2$ (i.e. included in $M_{LH}(f)$), below the first Nyquist zone. As we increase the quantizer resolution N, the first modulation side band at $f_0$ appears at a higher frequency. Therefore, multi-bit quantizers only have components in $M_{HF}(f)$.

In Fig.~\ref{fig:sidebands_vs_quantizer_size}, the side bands magnitude is proportional to the rest frequency $f_0$, as \eqref{eq:dt_spectrm} shows, therefore, it is also proportional to the number of bits N. Apparently this results in a paradox if we think in terms of a CTSD modulator. For instance, in a CTSD modulator with 3 bits and a given input, if we increase the quantizer resolution to 4 bits but we halve the input signal, the sequence obtained would have the same number of quantization levels. However, a detailed analysis of our PFM equivalent shows that also the PFM equivalent displays the same behaviour. In Fig.~\ref{fig:sidebands_vs_quantizer_size} we can observe that the increase in the side band magnitude with $f_0$ also corresponds with a higher attenuation of the sinc shaping pulse $H_{pulse}(f)$, which does not depend on $f_0$ or the number of bits N. Although apparently, a shift in the input DC may seem to produce different side band aliases after sampling, it can be proven that aliases are independent of the DC content.  

Let us consider the modulation side bands in $M(f)$. We can calculate the amplitude of the tones in each modulation side band after the sinc filter $H_{pulse}(f)$ 
(designated as $P(f)$) by using \eqref{eq:dt_spectrm}:
\begin{eqnarray}
  P(q f_0+r f_x) = \gamma_{q,r} \frac{\sin(\pi \left (q f_0+r f_x \right ))}{\pi \left (q f_0+r f_x \right )} \cdot f_0 \left (1+\frac{r f_x}{q f_0} \right ) 
  \label{eq:sidebands_sinc_compensation_long}
\end{eqnarray}
\begin{eqnarray}
  \gamma_{q,r} = J_r \left (q \frac{A k_d}{f_x} \right )
  \label{eq:gamma_bessel_etc}
\end{eqnarray}
In a multi-bit CTSD modulator, $f_0$ is an integer multiple of $f_s$, as a consequence, \eqref{eq:sidebands_sinc_compensation_long} can be simplified as follows:

\begin{equation}
 P(q f_0+r f_x) = \gamma_{q,r} \cdot \frac{1}{q} \frac{\sin(\pi r f_x )}{\pi} .
  \label{eq:sidebands_sinc_compensation_simplified}
\end{equation}

Thus, for modulators with different quantizer resolutions N, the amplitude of the tones in $P(f)$ for modulation side bands with same indexes q are identical and do not depend on the number of bits N. A similar mathematical analysis would show that also the phase of the tones in the modulation side bands matches, ensuring a total equivalence after sampling. Attempts to modify the SQNR of a CTSD modulator by intentionally shifting $f_0$, have been reported in several papers, for instance \cite{prefasi_npi}. 

Noise shaping in the PFM equivalent can be understood with the help of Fig.~\ref{fig:noise_source_comparison}(c). Observing Fig.~\ref{fig:noise_source_comparison}(c), we can see that $M_{LF}(f)$ passes the shaping pulse $H_{pulse}(f)$ only slightly attenuated, as it is within the main lobe of the sinc function, and will not be affected by aliasing. However, components $M_{HF}(f)$ will be subjected to aliasing. These high frequency components, after passing through the pulse shaper $H_{pulse}(f)$, will alias to the Nyquist band after sampling. Similarly as in~\cite{Gutierrez2018}, this combined effect can be modeled as the injection of a discrete-time aliasing component $m_{AL}[n]$ at the sampler, as shown in Fig.~\ref{fig:noise_source_comparison}(c). In prior work~\cite{Gutierrez2018}, the error component $M_{LF}(f)$ was denoted {\em the fundamental PFM encoding error}. In modulators of orders 2 and higher, the PFM is embedded in a control loop with extra integration in $L_{PFM}(s)$. Hence this low-frequency error component is also attenuated by the corresponding loop gain, which further increases the noise-shaping order of the error component. Since injection of $m_{AL}[n]$ happens at the sampler, we can define a discrete-time Alias Transfer Function $ATF(z)$. By inspection of Fig.~\ref{fig:noise_source_comparison}(c) it can be seen that:
  \begin{equation}
    ATF(z) =  \frac{1}{1+ L_{eq}(z)} \\
    \label{eq:ATF}
  \end{equation}
  where $L_{eq}(z)$ stands for the discrete time equivalent of the loop corresponding to the loop of Fig.~\ref{fig:noise_source_comparison}(c):
  \begin{equation}
    L_{eq}(z) = \left\{ L_{PFM}(s) \cdot \left(\frac{1-e^{-sT_s}}{s}\right)^2 \right\}^{\ast} \\
    \label{eq:L_eq}
  \end{equation}
Here $L_{PFM}(s)$ is the continuous loop filter with inner integrator, which forms the PFM, stripped off, see Fig.~\ref{fig:noise_source_comparison}(b). We recall that in \eqref{eq:L_eq}, function $H_{pulse}(s)$ appears squared because in Fig.~\ref{fig:noise_source_comparison}(c) there is one $H_{pulse}(s)$ function after the PFM modulator, cascaded with another $H_{pulse}(s)$ function representing the DAC. The impulse invariant transformation corresponding to the sampling \cite{fri04}, has been denoted as $\left\{ \cdot \right\}^{\ast}$. For instance, in the second order modulator of Fig.~\ref{fig:figure_2nd_single_bit} this function is:
 \begin{equation}
     ATF(z) =3 \cdot  \frac{1-z^{-2}}{3-2z^{-1}+z^{-2}}
\label{eq:ATF2nd}
 \end{equation}
Note that this transfer function is not the $NTF(z)$ of conventional Sigma-Delta theory and that the spectral density of its input is not a white noise, like in a uniform quantizer. Its meaning is to express the attenuation of the aliased modulation components produced in the PFM, which do not necessarily have a white spectrum.

The authors have identified the following mechanisms that can potentially produce in-band spurious tones and that can be explained using the PFM model:

\begin{itemize}
    \item I) Spurious tones due to re-injection of modulation side bands into the PFM modulator.
    \item II) Spurious tones coming from a modulation side band within the first Nyquist zone ($M_{LF}$).
    \item III) Spurious tones due to sub-sampling of high frequency modulation side band tones ($M_{HF}$).
\end{itemize}

Mechanism I) consists in the down conversion to the base band of high frequency side band tones when they re-enter in the PFM modulator, after being attenuated by the feedback loop $L_{PFM}(s)$. This mechanism is more relevant in VCO-based filters \cite{Talegaonkar, filtros_iscas} (which do not have a sampler in the loop) and has been analyzed in detail in \cite{vm_iscas2023}. For sake of conciseness we will not address this problem here. The other two mechanisms (II and III) will be discussed in the rest of this section. 

\subsection{Mechanism II: Spurious tones coming from a modulation side band within the first Nyquist zone}
To check how the PFM equivalent can predict the spectral behavior of a CTSD modulator, we have simulated the models of Fig.~\ref{fig:noise_source_comparison} in both equivalent versions of a second-order, single-bit modulator. In our simulation, the coefficients are $a_1=a_2=b_1=1, b_2=3/2$, $x_m=1$ and we have used the scaling of Fig.~\ref{fig:pfm_equivalence_2nd_single_bit}. Furthermore, we applied an input sine wave with an amplitude $x_{max}=0.24$ and an offset of $x_{DC}=0.25$ instead of $x_{DC}=1/2$ as would be required to center the input around $x_m$. This kind of signal produces a significant amount of in band harmonics in a second-order CTSD modulator, not attributable to circuit impairments. We are going to show how the proposed PFM model predicts these harmonic components. 

\begin{figure}
	\centering
	\includegraphics[width=\columnwidth,keepaspectratio]{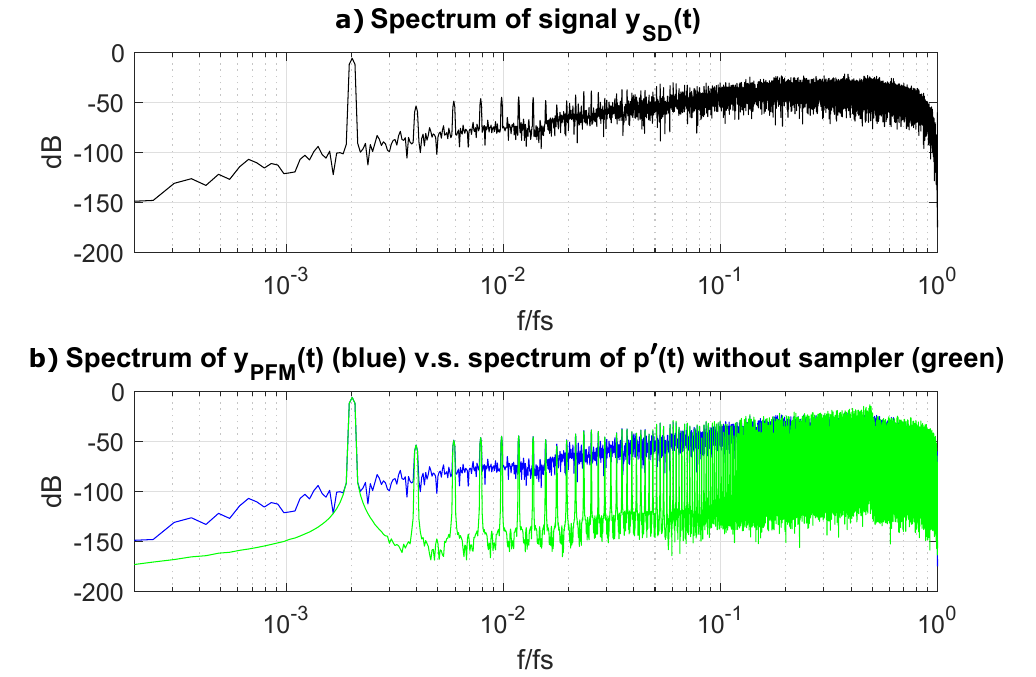}
	\caption{Analog spectra comparison of a real CTSD modulator and the predictions of the PFM model. (a) Spectrum of $y_{SD}(t)$. (b) Spectrum of $p'(t)$ (unsampled $y_{PFM}(t)$) compared to $y_{SD}(t)$.}  \label{fig:spectrum_comparison_2nd}
\end{figure}

Fig.~\ref{fig:spectrum_comparison_2nd} shows the analog Fourier transform of the different signals in Fig.~\ref{fig:noise_source_comparison}(b) (i.e. not the FFT of discrete sequences but the spectrum of the continuous-time signals) in the bandwidth between 0 and $f_s$. Fig.~ \ref{fig:spectrum_comparison_2nd}(a) shows the spectrum at the output of the equivalent CTSD modulator $y_{SD}(t)$. Here, spurious components (harmonics of the input signal), are clearly visible above the quantization noise floor. The presence of such spurious harmonic components can not be explained by the conventional white noise model, however, the PFM equivalent explains the presence of these components. By observing the single-bit case (N=1) of fig \ref{fig:sidebands_vs_quantizer_size} we see that lowering the input DC component, brings the first modulation side band at $f_0$ very close to the base band. We have done an additional simulation with a modified version of the system of Fig.~\ref{fig:noise_source_comparison}(b) removing the sampler, then $y_{PFM}(t)=p(t)$ and $e_{al}(t)=0$. For this modified system, we have designated the output signal  of the pulse shaper as $p^\prime(t)$. This signal $p^\prime(t)$ will fully exhibit the effect of the PFM modulation spurs but it will not be affected by any aliasing component (because the sampler was removed). Hence, when focusing on the Nyquist band (between 0 and $f_s/2$) all components that are present in  $p^\prime(t)$ will also be present in $p(t)$ of Fig.~\ref{fig:noise_source_comparison}(b). The resulting spectrum is shown in Fig.~\ref{fig:spectrum_comparison_2nd}(b) in green. The output of the PFM equivalent $y_{SD}(t)$, which is identical to $y_{SD}(t)$ in Fig.~\ref{fig:spectrum_comparison_2nd}(b), is  plotted for reference as a blue line in the figure. As can be seen, the harmonic components of the input signal which appear above the quantization error, are also present in the PFM equivalent modulator, indicating that the spurs have the same amplitude in both models.

\subsection{Mechanism III: Spurious tones due to sub-sampling of modulation side band tones}

In the previous subsection, we were focusing on the low-frequency component of the PFM spurs. Now we focus on the high frequency components which occur at frequencies which are beyond half the sampling frequency and therefore appear at the modulator output at different frequencies after being subsampled. The source of these spurious tones are the modulation side bands added by the PFM modulator when a tone at $f_x$ is applied at the input (signal $m(t)$ in Fig.~\ref{fig:noise_source_comparison}(b)). The modulation side bands, after appearing at the PFM modulator, pass through the pulse shaper $H_\textrm{pulse}$ and are sampled. Given the frequency location of each tone in the modulation side bands defined by \eqref{eq:dt_spectrm}, $f_{spur}(q,r)$:
\begin{equation}
f_{spur}(r,q) = q f_0 + r f_x
\label{eq:fspur}
\end{equation}
we can calculate the corresponding frequencies $f_a$ of the aliased tones in $m_{al}[n]$ (see Fig.\ref{fig:noise_source_comparison}(c)):
\begin{eqnarray}
f_a(r,q)= |f_{spur} - \text{nint}(f_{spur}/f_s) \cdot f_s | 
\label{eq:SpursInPFM}  
\end{eqnarray}

\begin{figure}[t]
	\centering
	\includegraphics[width=\columnwidth,keepaspectratio]{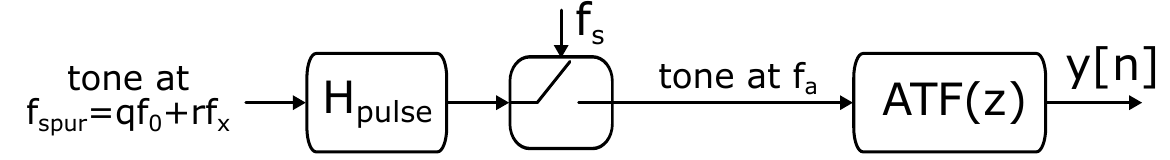}
	\caption{Signal flow diagram explaining the injection of limit cycles in a Sigma-Delta modulator.} \label{fig:SpurAliasingAndFiltering}
\end{figure}

In \eqref{eq:SpursInPFM}, $\text{nint}( \cdot )$ represents the nearest integer function using rounding half up rule and $| \cdot |$ the absolute value function. 
For a sinusoidal input, $m_{al}[n]$ is composed of the aliases of the infinite number of tones in the side bands of $M_{HF}(f)$. Actually, quantization noise is nothing but these aliases filtered by the $ATF(z)$ function defined in Section IV.A. The corresponding signal flow diagram is shown in Fig.~\ref{fig:SpurAliasingAndFiltering}.

\begin{figure}[t]
	\centering
	\includegraphics[width=\columnwidth,keepaspectratio]{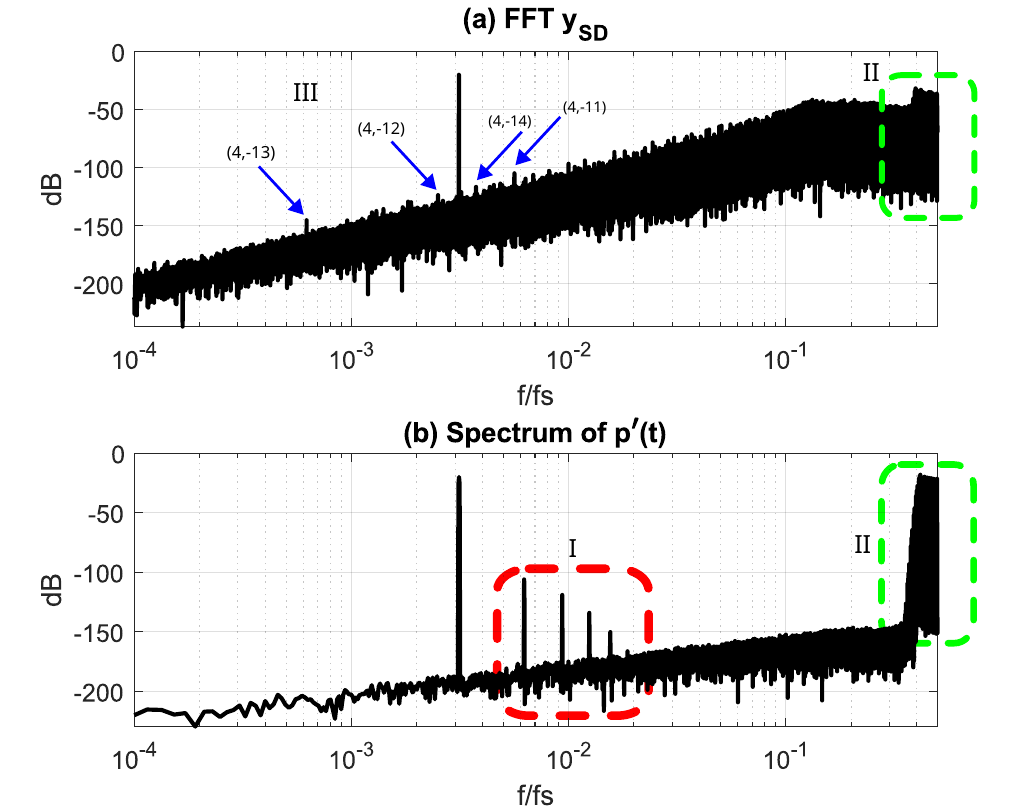}
	\caption{(a) FFT of $y_{SD}$ and (b) analog spectrum when sampler is removed. $q$ and $r$ are the coefficients that determine the frequency of the spur \eqref{eq:fspur}.} \label{fig:SpurAlias_HighFrequency_2ndOrder}
\end{figure}

Now two cases can be considered: either the modulation side band tones alias to harmonics of the input signal or they alias to non-harmonics frequencies. 

Harmonic distortion spurious tones appear when the DC component at the input $x_{DC}$, shifts the rest frequency of the PFM modulator $f_0$ to an integer multiple of the sampling frequency $f_s$ in \eqref{eq:SpursInPFM}. In this case, side bands around the $q^{th}$ harmonic of $f_0$ will be around $f_s$ and will alias to harmonics of $f_x$. Non harmonic distortion spurious tones appear when the DC component at the input $x_{DC}$, shifts the rest frequency of the PFM modulator to a non integer multiple of the sampling frequency $f_s$. In this case, spurious tones may appear at any location (even at frequencies lower than the input signal $f_x$) but are spaced in frequency by a distance $f_x$.

Fig.~\ref{fig:SpurAlias_HighFrequency_2ndOrder} shows a simulation of a single-bit second-order CTSD modulator with the following parameters $a_1=a_2=b_1=b_2=1, f_s=1, x_{DC}=0.51$ and a $-14~\text{dB}_{\text{FS}}$ input tone. Fig.~\ref{fig:SpurAlias_HighFrequency_2ndOrder}(a) shows the FFT of $y_{SD}$ where we have marked in blue the indexes $(q,r)$ of several tones $f_a$ coming from the aliasing of the side bands, identified using \eqref{eq:SpursInPFM}. After aliasing, these tones do not fall at harmonics of the input signal, due to $x_{DC}$ pushing $f_0$ to a non-rational factor of $f_s$. Our frequency prediction accurately identifies the spurious tones, visible in the FFT of the time domain simulation. The amplitude of the spurious tones was estimated using \eqref{eq:ATF2nd} and \eqref{eq:dt_spectrm}. For these tones, the estimation error is up to 2 dB, but it must be considered that amplitude prediction is compromised by the truncation of the infinite series in \eqref{eq:dt_spectrm}. 

In Fig.~\ref{fig:SpurAlias_HighFrequency_2ndOrder}(a), we can also separate the effect of the spurious tones coming from the Nyquist band by removing the sampler in the loop, like in Fig.~\ref{fig:spectrum_comparison_2nd}(b). Fig.~\ref{fig:SpurAlias_HighFrequency_2ndOrder}(b) displays the spectrum of $p^\prime(t)$ which shows the spurs due to Mechanism I and II previously described. The second and third-harmonic (red) are due to the re-injection of modulation side bands into the PFM modulator \cite{medina_iscas} (Mechanism I) and should be absent in the Sigma-Delta modulator due to the effect of the sampler. Since $x_{DC}=0.51$, the first side band centered at $f_0$ (green dashed box) is near half of $f_s$ (Mechanism II) which is also visible in the FFT of $y_{SD}$ shown in Fig.~\ref{fig:SpurAlias_HighFrequency_2ndOrder}(a).

\section{Conditions for PFM equivalence of a Sigma-Delta modulator under quantizer overload}

It is known that the dynamic range of single-bit CTSD modulators has some limitations close to full-scale, in special the second-order modulator \cite{ardalan}. The PFM equivalent is also affected by such limitations. This Section is devoted to analyze this fact, revealing the conditions under which the PFM equivalence is valid and giving further insight into the reasons for the dynamic range limitation and even the implications in the stability of a CTSD modulator. 

Going back to the definition of the PFM equivalent of a first-order single-bit Sigma-Delta modulator of Fig.~\ref{fig:fig1_VM}, we stated that the equivalent of Fig.~\ref{fig:fig1_VM}(c) is valid for both single-bit and multi-bit modulators. On the contrary, the Sigma-Delta modulator of Fig.~\ref{fig:fig1_VM}(a) always produces codes either 0 or 1 regardless of the input. The PFM equivalent of Fig.~\ref{fig:fig1_VM}(c) will be valid and, according to Appendix A, identical to the Sigma-Delta modulator as long as the PFM equivalent does not generate more than two levels. The reason why the system of Fig.~\ref{fig:fig1_VM}(c) may produce more than two levels is because two delta pulses in $d(t)$ occur in less than the shaping pulse $H_{pulse}$ duration $T_s$. If this happens, one pulse appears at $p(t)$ before the previous one has vanished, giving rise to a multilevel signal. The mathematical condition for this to happen can be deducted from \eqref{eq:pfm_recurrence}. If there exist any $t_k$ for which input $x(t)$ complies with the following condition, the PFM equivalence will not hold:

\begin{eqnarray}
\int_{t_{k-1}} ^{t_k} x(t) dt-x_m \cdot T_s = 0 \nonumber \\
\exists t_k \mid  (t_{k}-t_{k-1}) < T_s
 \label{eq:pfm_saturation}
\end{eqnarray}

For DC inputs, this condition translates into the maximum DC input $x_{DC} \leq x_{m}$ if $T_m=T_s$ in \eqref{eq:pfm_dc_input}. This bound can approximately be used as well for highly oversampled signals as those at the input of a first-order Sigma-Delta modulator. 
However, when we have a high order Sigma-Delta modulator like in Fig.~\ref{fig:noise_source_comparison}(b), the PFM modulator in the PFM equivalent is at the end of the loop filter, and depending on the scaling of state variables, the input signal to the PFM modulator can have a very different amplitude than the input signal. Moreover, the signal at the end of the loop filter is highly incorrelated and the DC approximation of \eqref{eq:pfm_saturation} might no be so accurate. 

\begin{figure}[t]
	\centering
	\includegraphics[width=\columnwidth,keepaspectratio]{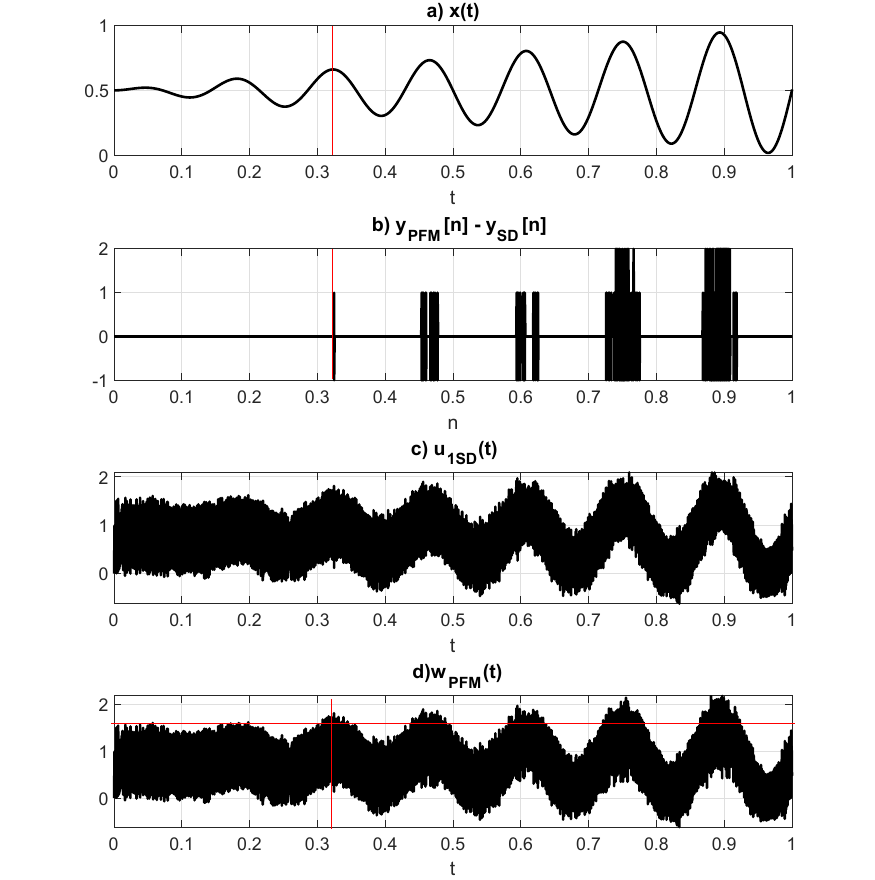}
	\caption{Quantizer overload and PFM modulator overload points in a single-bit second-order modulator.} 
    \label{fig:figure_2nd_ramp}
\end{figure}

We are going to analyze next the input range of the single-bit, second-order modulator described in Section III. In Fig.~\ref{fig:figure_2nd_ramp} we have represented a simulation of both a single-bit second-order Sigma-Delta modulator and its PFM equivalent system, as depicted in Fig.~\ref{fig:pfm_equivalence_2nd_single_bit} with the standard coefficients ($a_1=a_2=b_1=1, b_2=3/2, x_m=1$). In this simulation, a slow sine wave with growing amplitude between 0 and 1/2 and an offset $x_{DC}=1/2$ is applied to the input (see Fig.~\ref{fig:figure_2nd_ramp}(a)). In Fig.~\ref{fig:figure_2nd_ramp}(b) we have plotted the difference between the output of the PFM equivalent system $y_{PFM}$ and the output of the Sigma-Delta modulator $y_{SD}$. We can see that the two systems produce the same signal up to some point where the equivalence breaks (see red vertical line). In Fig.~\ref{fig:figure_2nd_ramp}(c), we have plotted state variable $u_{1SD}(t)$ and in Fig.~\ref{fig:figure_2nd_ramp}(d) the input to the PFM modulator, $w_{PFM}(t)$. According to \eqref{eq:pfm_saturation}, the point at which the PFM modulator jumps from single-bit to multi-bit should happen at $w_{PFM}>3/2$ if $b_2=3/2, a_2=1$, and $f_s=1$. In Fig.~\ref{fig:figure_2nd_ramp}(d) the value of $w_{PFM}$ approximates this value close to the first sample at which the equivalence fails. The value cannot be exactly that predicted by \eqref{eq:pfm_saturation} because this equation is only valid for DC inputs and state variable $w_{PFM}$ has energy at high frequency, as a difference from $x(t)$, therefore, the saturation point is dependent on the input signal as well. However, this saturation point in $w_{PFM}$ corresponds to a small input signal amplitude, as represented in Fig.~\ref{fig:figure_2nd_ramp}(a) of approximately sine amplitude $x=0.17$ or $-9~\text{dB}_{\text{FS}}$.

\begin{figure}[t]
	\centering
	\includegraphics[width=\columnwidth,keepaspectratio]{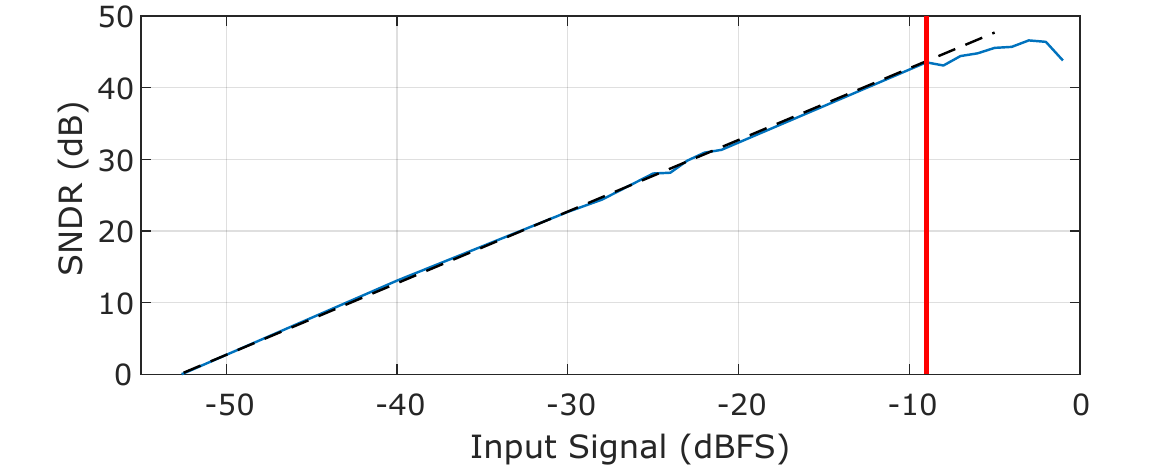}
	\caption{Dynamic range of a single-bit second-order CTSD modulator with $f_x=f_s/2400$ and $OSR = 20$. The SNDR becomes dominated by distortion above the point where PFM coding limit is exceeded.} 
    \label{fig:figure_2nd_DR_drop}
\end{figure}

\begin{figure}[t]
	\centering
	\includegraphics[width=\columnwidth,keepaspectratio]{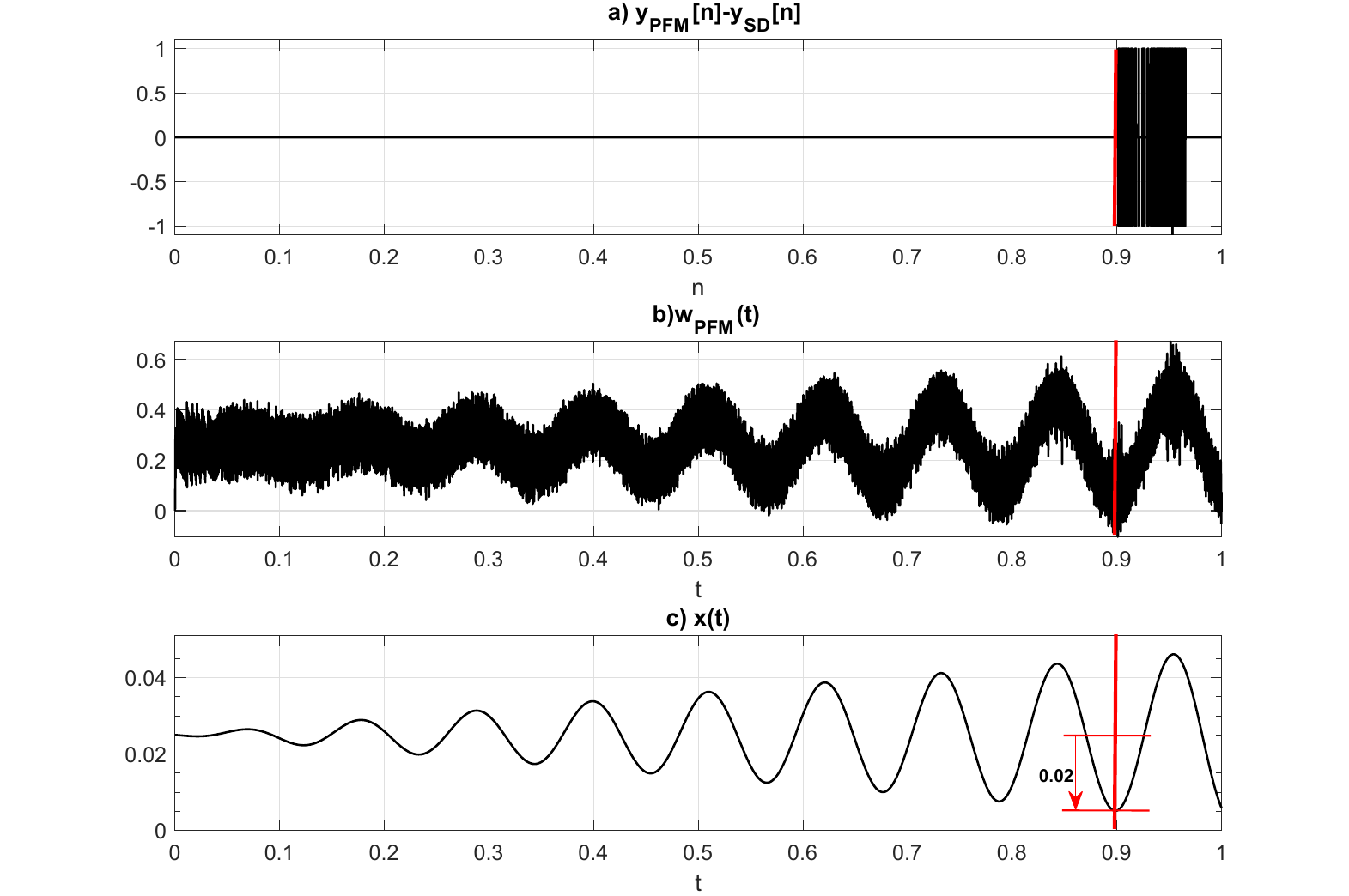}
	\caption{Quantizer overload and PFM modulator overload points in a single-bit 3rd order modulator.} 
    \label{fig:figure_3rd_ramp}
\end{figure}

When the input of a CTSD modulator drives the internal PFM modulator of its equivalent model out of the bounds of the CTSD modulator quantizer, we will say that we have \textit{exceeded the PFM coding limit} for that resolution. The Sigma-Delta modulator may still work for some input range before getting unstable but the quantization error may not be represented by PFM side bands plus aliased components. Moreover, from this point, the Signal-to-Noise and Distortion Ratio (SNDR) of the CTSD modulator will be impaired. This behaviour is shown in Fig.~\ref{fig:figure_2nd_DR_drop}, which represents the dynamic range plot of the modulator used in Fig.~\ref{fig:pfm_equivalence_2nd_single_bit}. Unfortunately and in spite of being one of the most successful hardware implementations, single-bit second-order modulators are especially prone to this problem. 
Most techniques devoted to stabilize Sigma-Delta modulators of high order \cite{schreier_ct}, at the end result in a scaling of state variables to avoid the PFM coding limit overload. For instance, we are going to repeat the analysis of Fig.~\ref{fig:figure_2nd_ramp} with the third-order modulator that we used as example in Section III.C. According to the coefficient scaling of Section III.C, the input range to operate as a single-bit modulator is $0 \le x(t) \le b_1$, where $b_1=0.05$. In the simulation, we have used a linearly growing sine wave with a DC component $x_{DC}=b_1/2$. Fig.~\ref{fig:figure_3rd_ramp}(a) shows the difference between the Sigma-Delta and PFM model outputs, which match sample by sample until the input amplitude corresponds to 
a sine amplitude of 0.02 (see red line), very close to the full scale value $b_1/2=0.025$. At this point, the dynamic range of the CTSD modulator falls sharply and the PFM equivalent turns multi-bit. Fig.~\ref{fig:figure_3rd_ramp}(b) shows state variable $w_{PFM}$ which is always within the operating range of the PFM modulator (0 to $x_m=1$) until the modulator equivalence breaks because $w_{PFM}$ goes out of bounds. Finally Fig.~\ref{fig:figure_3rd_ramp}(c) reproduces the input signal. 

It must be noted that the same quantizer overload affects multi-bit Sigma-Delta modulators, except that quantizer saturation only happens for the last LSB of the quantizer range and hence, the dynamic range limitation is down-scaled by the number of bits.

\section{Conclusion}

The classical modelling of a Sigma-Delta modulator mimics its circuit implementation. However, and as a main conclusion, we show that a better understanding can be achieved by defining a reference system (the PFM equivalent), at a higher abstraction level and with a lesser relationship to the circuit. For instance, early Sigma-Delta modulators employed a flash ADC and therefore, a uniform quantizer was considered an essential piece of any modulator. In our vision of Sigma-Delta modulation, uniform quantization is rather a consequence of the reference system construction, whose relationship to the circuit is not straightforward. 
Our work is more a collection of observations around the PFM equivalent than a closed mathematical theory. These observations provide a cause-effect relationship for most of the nonlinear phenomena observed in Sigma-Delta modulators, which were previously described by simulation or statistics. Nevertheless, we consider that our work is complementary to the classical understanding of Sigma-Delta modulators. Modelling based on noise transfer functions and white quantization noise has proven to be very efficient for practical design.

As a summary of our contributions in this paper, we count on the sampling invariance as the enabling concept for the rest of the findings. Then, we propose two concepts, the notion that multi-bit and single-bit modulators are the same system under different parameters and the relationship between the PFM equivalent and the quantizer overload of a Sigma-Delta modulator. Also, the origin of spurious tones is revealed as being either subsampled or Nyquist band PFM modulation components. There are still a number of topics to be studied, such as application to MASH modulators, analysis of loop delay or other non CIFB structures, to be covered in future works.

\section* {Acknowledgments}
This paper was supported by project PID2020-118804RB-I00 of the Spanish Agency of Research (AEI) and by FWO-Vlaanderen.

\begin{appendices}
\section{Proof of PFM and first-order CTSD Modulator equivalence}
In this appendix, the mathematical proof for the equivalence of a first-order CTSD modulator and a sampled PFM system is demonstrated. We will use the complete induction method and will prove that if for the same input, the samples of the CTSD modulator and the samples of its PFM equivalent match for all samples $k<n$, they will also match for sample $n$. For this purpose, the CTSD modulator is composed of a multi-bit quantizer with quantization step $\Delta$. 
If $0 \leq x(t) \leq \Delta$, then, the CTSD modulator behaves as a single-bit system. 

\begin{figure}[t]
	\centering
	\includegraphics[width=\columnwidth,keepaspectratio]{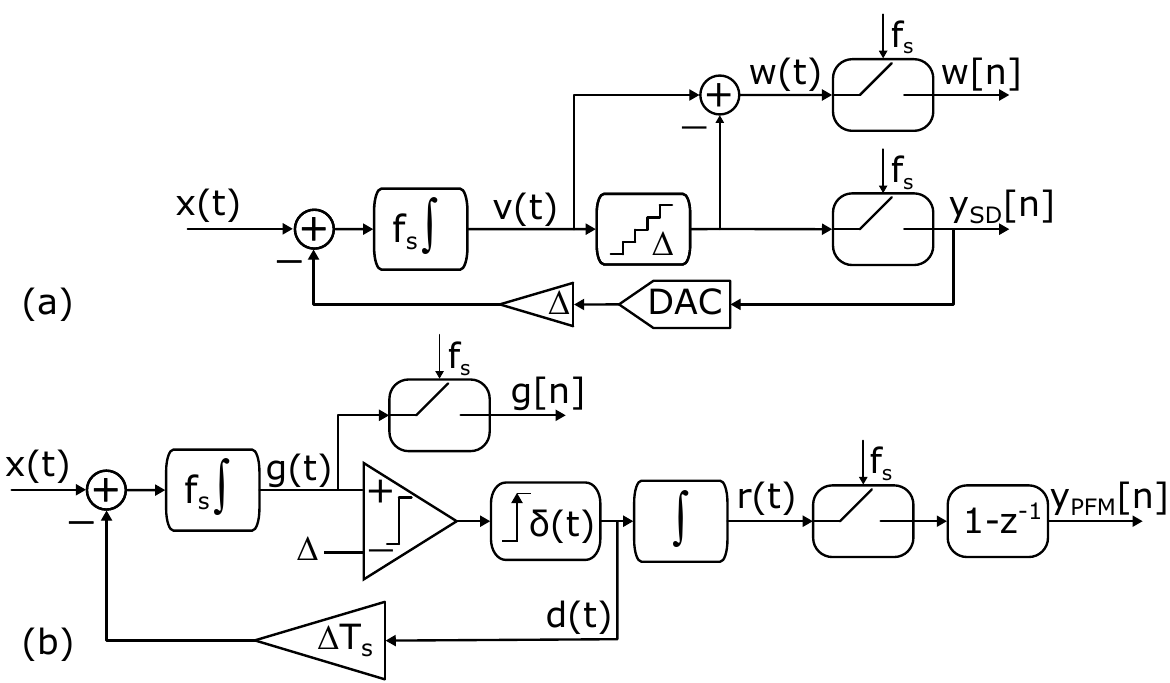}
	\caption{Models for sampling equivalence proof.
	} \label{fig:fig1appendix_mathematicalEquivalence}
\end{figure}

Fig.~\ref{fig:fig1appendix_mathematicalEquivalence} shows a model of a first-order CTSD modulator and a PFM modulator with some auxiliary signals that we will make use of. We will assume that the three integrators in Fig.~\ref{fig:fig1appendix_mathematicalEquivalence} are at rest at $t=0$. Let $w(t)$ be the difference between $v(t)$ and its quantized version $Q[v(t)]$, and let $g(t)$ be the state variable of the PFM modulator. Note that the integrator producing $r(t)$ followed by a sampler and a first-difference function is equivalent to the block $h_{pulse}(t)$ followed by a sampler used in the previous sections.  Based on these definitions, we can build the following hypothesis:
\begin{eqnarray}
	\text{if } \forall k<n, y_{SD}[k]=y_{PFM}[k] \implies w[n] = g[n], \\
	\text{if } w[n] = g[n] \implies y_{SD}[n] = y_{PFM}[n],
	\label{eq:hyphotesis}
\end{eqnarray} 

The equations that describe signal $w(t)$ at instant $nT_s$ are as follows:
\begin{eqnarray}
	w(n T_s)=v(n T_s)-Q[v(n T_s)] = v(n T_s) - \Delta y_{SD}[n] 
	\label{eq:SD_w1}
\end{eqnarray}
\begin{equation}
    \begin{split}
        v&(n T_s) = \frac{1}{T_s} \int_{-\infty}^{n T_s} \biggl(x(\tau)- \Delta \sum_{k=-\infty}^{n-1}y_{SD}[k] \cdot c(\tau) \biggr) \,d\tau \\
        &=\frac{1}{T_s} \int_{(n-1)T_s}^{n T_s} \biggl(x(\tau) - \Delta y_{SD}[n-1] \biggr) \,d \tau +v((n-1)T_s) \\
        &=\frac{1}{T_s} \int_{(n-1) Ts}^{n T_s} x(\tau) \,d\tau + v((n-1)T_s) - \Delta y_{SD}[n-1] 
    \end{split}
    \label{eq:SD_u1}
\end{equation}
In \eqref{eq:SD_u1}, $c(\tau) = u(\tau-k T_s)-u(\tau-(k-1) T_s)$, where $u( \cdot )$ is the Heaviside unit step function. Operating with \eqref{eq:SD_w1} and \eqref{eq:SD_u1}, we obtain the following:
\begin{equation}
    \begin{split}
        w(n T_s) = &\frac{1}{T_s} \int_{(n-1) Ts}^{n T_s} x(\tau) \,d\tau -\Delta y_{SD}[n] \\
        &- \Delta y_{SD}[n-1]+v((n-1)T_s) \\
        =& \int_{(n-1) Ts}^{n T_s}x(\tau) \,d\tau + w((n-1)T_s)- \Delta y_{SD}[n]
    \end{split}
\end{equation}

Similarly, the equations to describe the signal $g(t)$ in the equivalent PFM system shown in Fig.~\ref{fig:fig1appendix_mathematicalEquivalence}(b) are as follows:
\begin{equation}
    \begin{split}
        g(t)=\frac{1}{T_s} \int_{-\infty}^{t} \biggl(x(\tau)-\Delta T_s d(\tau) \biggr) \,d\tau
    \end{split}
    \label{eq:SD_g1}
\end{equation}
\begin{equation}
    r(t)=\int_{-\infty}^{n T_s} d(\tau) \,d \tau = \int_{-\infty}^{n T_s} \sum_{j=-\infty}^{\infty} \delta(\tau-t_j) \, d \tau
    \label{eq:SD_r1}
\end{equation}
\begin{equation}
    t_j \, \mid \, \frac{1}{T_s} \int_{t_{j-1}}^{t_j} x(\tau) \, d \tau = \Delta
    \label{eq:SD_tj1}
\end{equation}
\begin{equation}
    y_{PFM}[n]=\int_{(n-1) T_s}^{n T_s} d(\tau) \, d \tau=r[n]-r[n-1]
    \label{eq:SD_yPFM1}
\end{equation}
If we now consider the instant $n T_s$:
\begin{equation}
    \begin{split}
        g(n T_s) =& \int_{(n-1) T_s}^{n T_s} x(\tau) \,d \tau + g((n-1)T_s) \\
        &- \Delta(r(nT_s)-r((n-1)T_s) \\
        =&\int_{(n-1) T_s}^{n T_s} x(\tau) \,d \tau + g((n-1)T_s)- \Delta y_{PFM}[n]
    \end{split}
    \label{eq:SD_g1_nTs}
\end{equation}

If $w[k]=g[k]$ for all $ k<n$, then:
\begin{equation}
    g(nT_s)=\int_{(n-1) T_s}^{n T_s} x(\tau) \,d \tau + w((n-1)T_s)- \Delta y_{PFM}[n]    
    \label{eq:SD_g1_wn-1}
\end{equation}
From \eqref{eq:SD_w1}, \eqref{eq:SD_u1} and \eqref{eq:SD_g1_nTs} we obtain the following:
\begin{equation}
    g(n T_s)- \Delta y_{PFM}[n] = w(n T_s)-\Delta y_{SD}[n]
\end{equation}
Now, by inspection of  Fig.~\ref{fig:fig1appendix_mathematicalEquivalence} we see that $g[n]$ and $w[n]$ are always positive and bounded by $\Delta$. Moreover, $y_{SD}[n]$ and $y_{PFM}[n]$ must be positive integers. Then:
\begin{eqnarray}
    g[n]<\Delta \, ,w[n]<\Delta  \nonumber \\
    y_{SD}[n] \in \mathbb{N}  \, , y_{PFM}[n] \in \mathbb{N}  \nonumber \\
    \implies y_{SD}[n]=y_{PFM}[n]\, ,g[n]=w[n]
    \label{eq:eureka}
\end{eqnarray}
To close the complete induction proof, we recall that the three integrators in Fig.~\ref{fig:fig1appendix_mathematicalEquivalence} are at rest at $t=0$, therefore $v(0)=0$, $r(0)=0$ and $g(0)=0$. As a consequence, $y_{SD}[0]=0$ and $y_{PFM}[0]=0$ which shows that the hypothesis holds for any index $0,1 \ldots n$.

\section{Extension to higher order modulators: Generic methodology}

\begin{figure}
	\centering
	\includegraphics[width=\columnwidth,keepaspectratio]{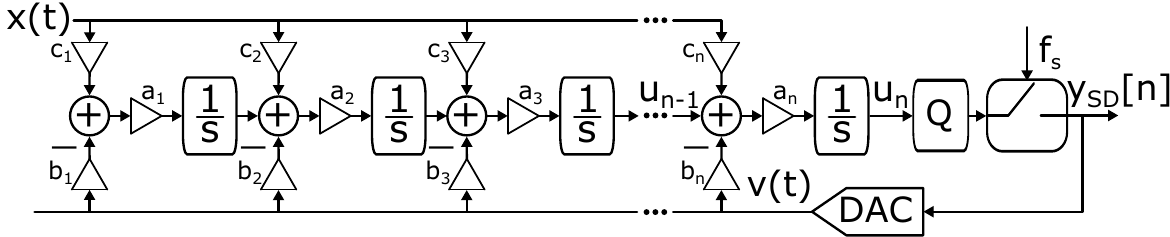}
	\caption{Generic CTSD modulator with a CIFB structure with feedforward.}  \label{fig:generic_CTDSM_CIFB_feedforward}
\end{figure}

We show in this appendix how to transform any CTSD modulator with the CIFB structure \cite{Schreier2017understanding} into its PFM equivalent. For sake of conciseness we will assume there is no feed-forward before the quantizer, but we will consider feed-forward coefficients into the other integrators. Extension to a wider class of modulators will be attempted in future works. Let us consider the $n^{th}$-order CTSD modulator shown in Fig.~\ref{fig:generic_CTDSM_CIFB_feedforward}, where for simplicity $f_s=1$. The equivalent PFM-based model will depend on whether the quantizer is single-bit or multi-bit as it was discussed in previous sections. These equivalent models will also present different transfer functions depending on the quantizer. The equivalent input transfer function, $L_{FS}'(s)$, and feedback transfer function, $L_{PFM}(s)$ (see Fig.~\ref{fig:noise_source_comparison}) will differ for multi-bit and single-bit cases as two models were identified for each case. The transfer functions for the single-bit equivalent system are as follows:
\begin{equation}
    L_{PFM}(s)= \frac{1}{b_n s^{n-1}} \sum_{i=1}^{n-1} b_i s^{i-1} \prod_{j=i}^{n-1} a_j \, , \, n > 1
    	\label{eq:HFd_singlebit}
\end{equation}

\begin{equation}
    L_{FS}'(s)= \frac{1}{b_n} \left( c_n + \frac{1}{s^{n-1}} \sum_{i=1}^{n-1} c_i s^{i-1} \prod_{j=i}^{n-1} a_j \right)
    	\label{eq:HFs_singlebit}
\end{equation}
If the order of the system is $n=1$, $L_{PFM}(s)=0$. Since for single-bit CTSD modulators the gain of the last integrator is irrelevant, the coefficient $a_n$ won't appear in $L_{PFM}(s)$ nor $L_{FS}'(s)$.

For a multi-bit system the transfer functions are as follows:
\begin{equation}
    L_{PFM}(s)= \beta + \frac{1}{s^{n-1}} \sum_{i=1}^{n-1} b_i s^{i-1} \prod_{j=i}^{n} a_j 
    	\label{eq:HFs_multibit}
\end{equation}
\begin{equation}
    L_{FS}'(s)= a_n c_n + \frac{1}{s^{n-1}} \sum_{i=1}^{n-1} c_i s^{i-1} \prod_{j=i}^{n} a_j
    	\label{eq:HFd_multibit}
\end{equation}
Coefficient $\beta=a_n \cdot b_n -1$. This coefficient was described in Section \ref{section_equivalence}. In this case, the coefficient $a_n$ is relevant and thus appears in both $L_{PFM}(s)$ and $L_{FS}'(s)$. 

\end{appendices}

\bibliography{referencias}
\bibliographystyle{IEEEtran}

\begin{IEEEbiography}[{\includegraphics[width=1in,height=1.25in,clip,keepaspectratio]{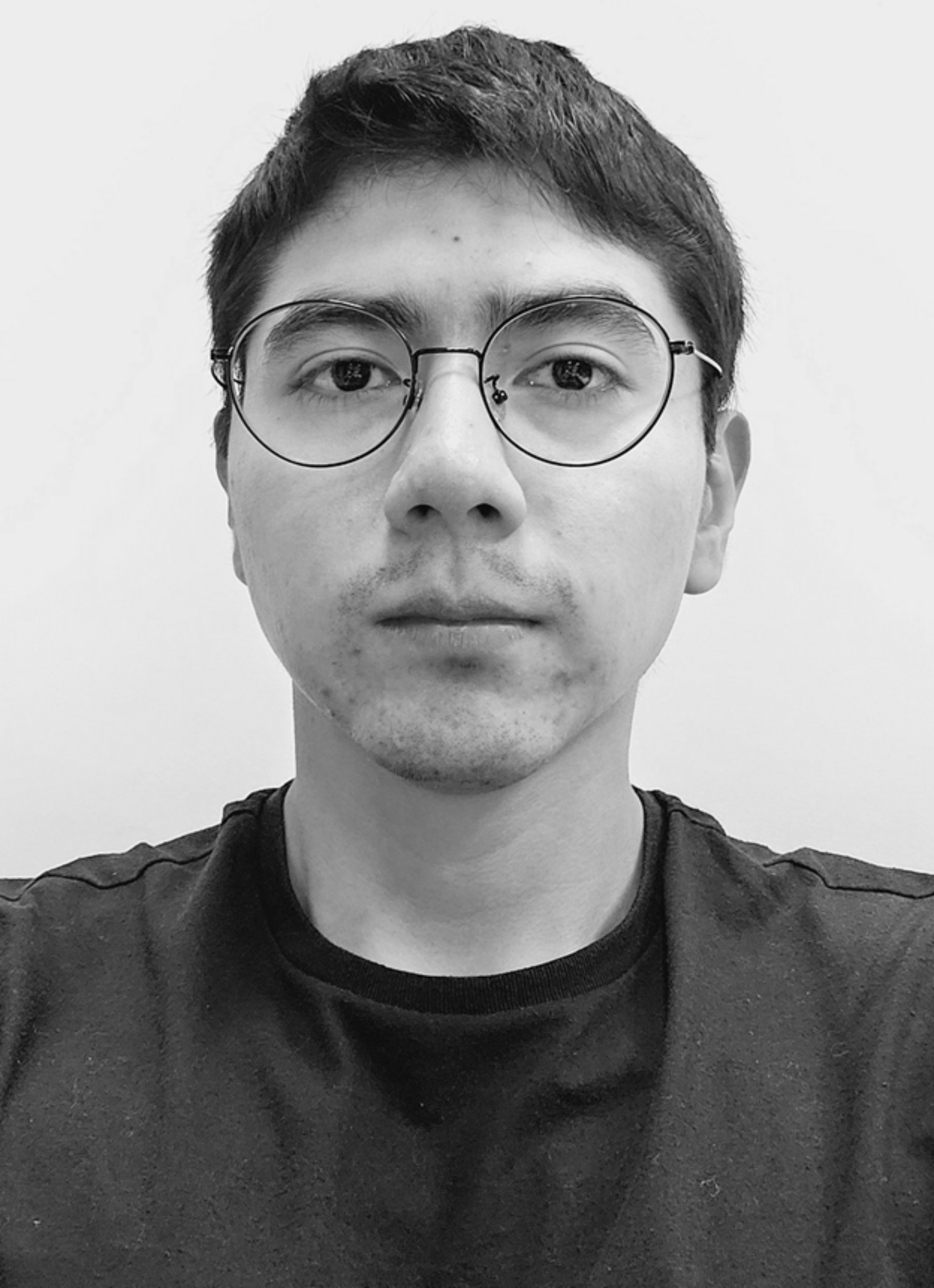}}]{Victor Medina} received the B.Sc. and M.Sc. degrees in Electronic Engineering from Carlos III University of Madrid, Spain, in 2018 and 2019, where he is currently pursuing the Ph.D. degree. In 2022, he did a four-month internship at Infineon Technologies, Villach, Austria. His current research interests include mixed-signal integrated circuit design and time-encoded systems theory.
\end{IEEEbiography}

\begin{IEEEbiography}[{\includegraphics[width=1in,height=1.25in,clip,keepaspectratio]{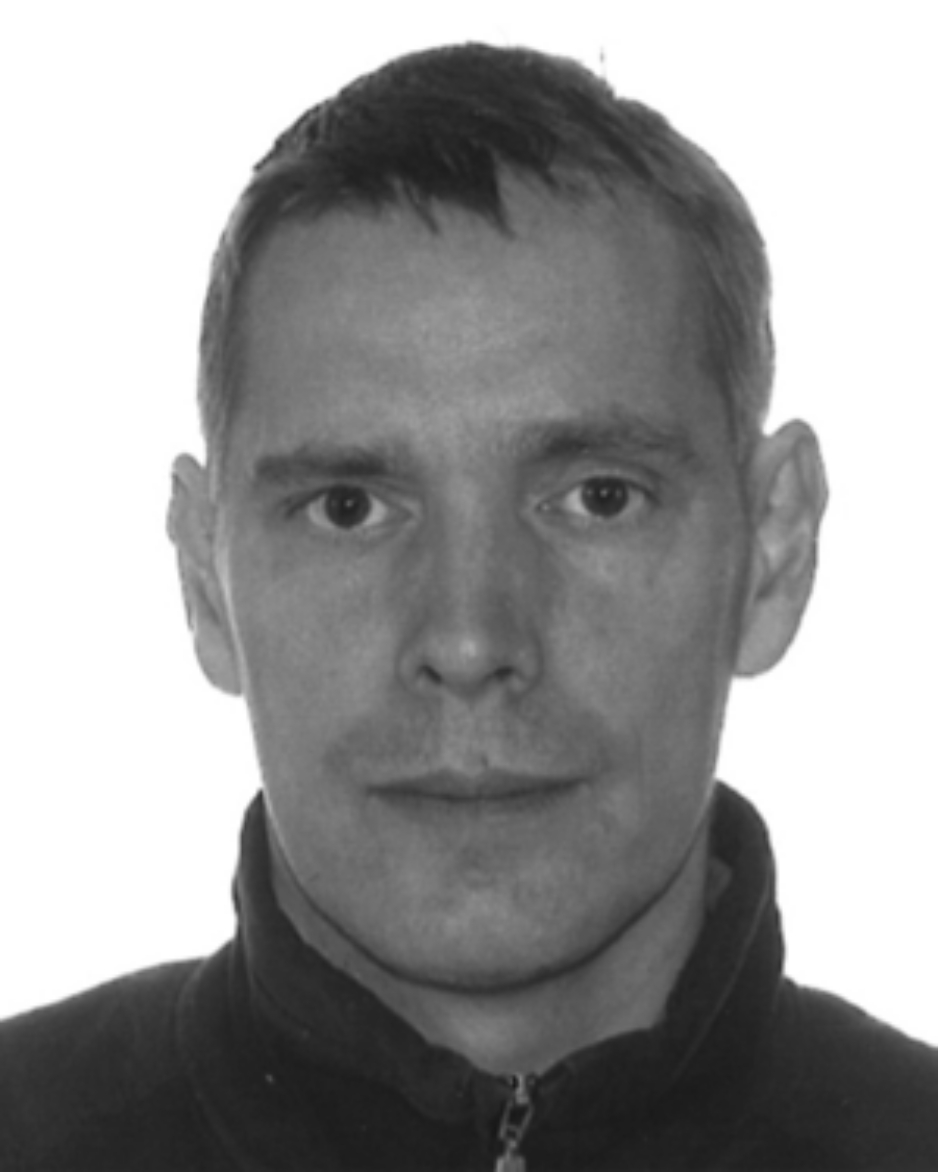}}]{Pieter Rombouts} received the Ir. and Ph.D. degrees from Ghent University in 1994 and 2000, respectively. He is currently a Professor of analog electronics at Ghent University. His research interests include signal processing, circuits and systems theory, analog circuit design, and sensor systems. The main focus of his research has been on A/D and D/A conversion. In the past, he has served as an Associate Editor for the IEEE Transactions on Circuits and Systems-I, the IEEE Transactions on Circuits and Systems-II and Electronics Letters.
\end{IEEEbiography}

\begin{IEEEbiography}[{\includegraphics[width=1in,height=1.25in,clip,keepaspectratio]{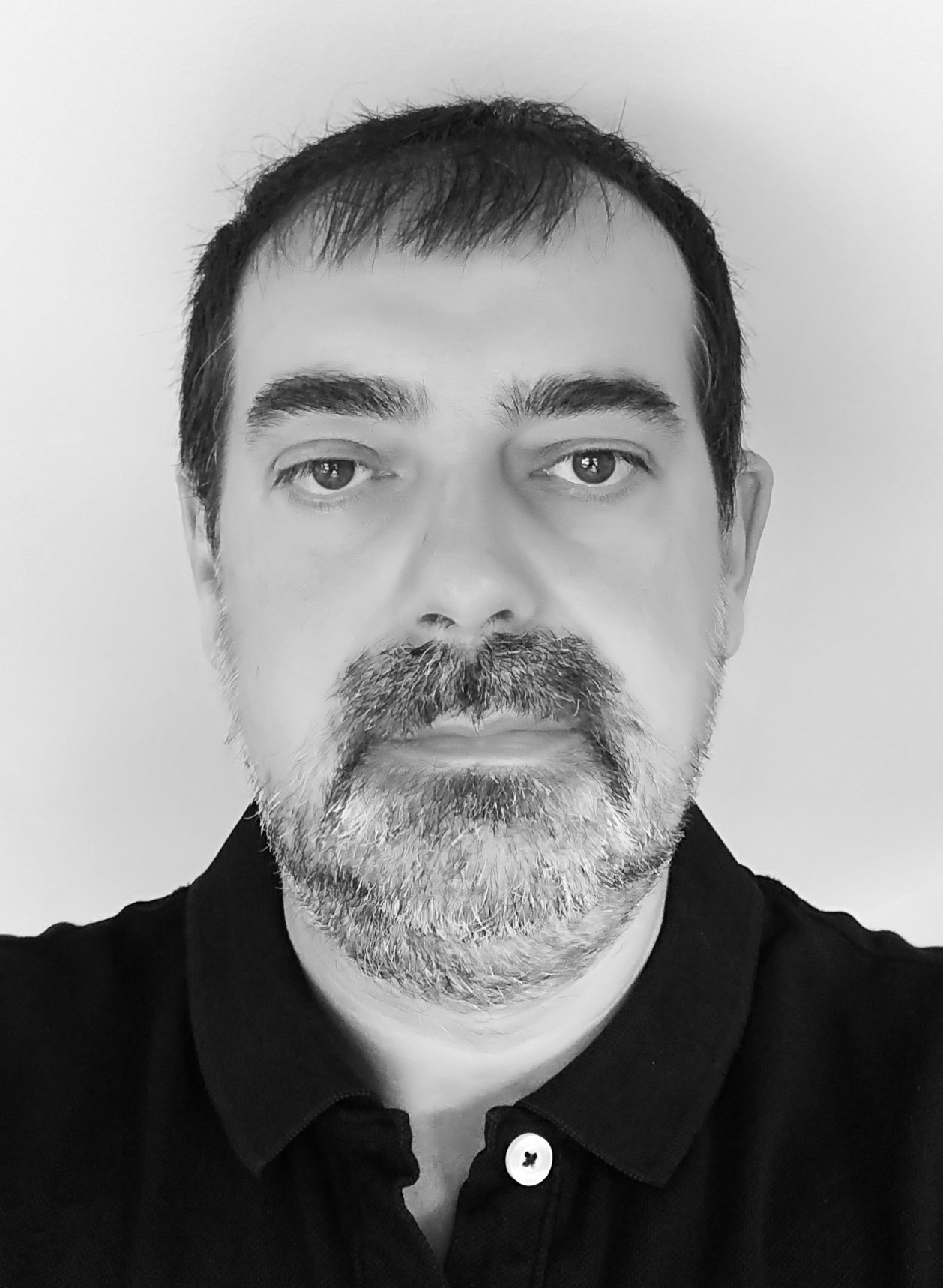}}]{Luis Hernandez-Corporales} received a MS (‘89) and PhD (’95) degrees in Telecommunication Engineering from Polytechnic University of Madrid. He did a postdoctoral stay at the ECE dept. of Oregon State University in 1996 and Analog Devices in Willmington, USA in 1997. In 1998 he joined University Carlos III of Madrid where he is currently Full Professor in the Electronic Technology Department and leads the mixed signal research group. He has been department head and PhD program director. In 2009 he did a sabbatical stay at IMEC, Leuven, Belgium. He has coauthored three books, over 170 papers and holds 25 patents. He is a member of the IEEE-CAS ASPTC committee and has been associate editor of IEEE Transactions on Circuits and systems I and II for 9 years. His topics of interest are analog microelectronics, Sigma-Delta modulation  time-encoded data converters and neural networks.
\end{IEEEbiography}

\end{document}